\documentclass[longauth]{aa}  

\usepackage{graphicx}
\usepackage{txfonts}
\usepackage{hyperref}
\usepackage{booktabs}
\usepackage{academicons}
\usepackage{xcolor}
\usepackage{fontawesome}
\usepackage{xcolor}
\usepackage[normalem]{ulem}
\usepackage{comment}
\usepackage{placeins}

\usepackage{chemformula}

\DeclareSymbolFont{UPM}{U}{eur}{m}{n}
\DeclareMathSymbol{\umu}{0}{UPM}{"16}
\let\oldumu=\umu
\renewcommand\umu{\ifmmode\oldumu\else$\oldumu$\fi}

\begin{document}

\title{CHEOPS photometry from 2024 reveals a reversal in the transit-timing variations of AU\,Mic\,c\thanks{This paper uses CHEOPS data observed as part of the Guaranteed Time Observation (GTO) programme CH\_PR140071.}$^{,}$\thanks{\email{zgarai@gothard.hu; zgarai@ta3.sk}}}
\subtitle{}

\author{
Z.~Garai\inst{\ref{inst:2},\ref{inst:3}}\,$^{\href{https://orcid.org/0000-0001-9483-2016}{\protect\includegraphics[height=0.19cm]{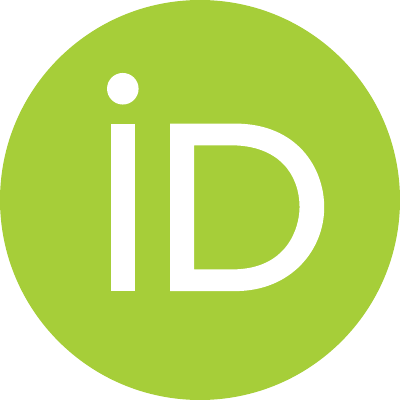}}}$, 
Gy.~M.~Szabó\inst{\ref{inst:2}}\,$^{\href{https://orcid.org/0000-0002-0606-7930}{\protect\includegraphics[height=0.19cm]{figures/orcid.pdf}}}$, 
D.~Gandolfi\inst{\ref{inst:4}}\,$^{\href{https://orcid.org/0000-0001-8627-9628}{\protect\includegraphics[height=0.19cm]{figures/orcid.pdf}}}$, 
A.~Brandeker\inst{\ref{inst:5}}\,$^{\href{https://orcid.org/0000-0002-7201-7536}{\protect\includegraphics[height=0.19cm]{figures/orcid.pdf}}}$, 
A.~Bonfanti\inst{\ref{inst:12}}\,$^{\href{https://orcid.org/0000-0002-1916-5935}{\protect\includegraphics[height=0.19cm]{figures/orcid.pdf}}}$, 
N.~Billot\inst{\ref{inst:6}}\,$^{\href{https://orcid.org/0000-0003-3429-3836}{\protect\includegraphics[height=0.19cm]{figures/orcid.pdf}}}$, 
W.~Benz\inst{\ref{inst:7},\ref{inst:8}}\,$^{\href{https://orcid.org/0000-0001-7896-6479}{\protect\includegraphics[height=0.19cm]{figures/orcid.pdf}}}$, 
A.~Heitzmann\inst{\ref{inst:6}}\,$^{\href{https://orcid.org/0000-0002-8091-7526}{\protect\includegraphics[height=0.19cm]{figures/orcid.pdf}}}$, 
G.~Olofsson\inst{\ref{inst:5}}\,$^{\href{https://orcid.org/0000-0003-3747-7120}{\protect\includegraphics[height=0.19cm]{figures/orcid.pdf}}}$, 
L.~Kriskovics\inst{\ref{inst:9},\ref{inst:10}}, 
Á.~Boldog\inst{\ref{inst:9},\ref{inst:10}}, 
L.~Borsato\inst{\ref{inst:11}}\,$^{\href{https://orcid.org/0000-0003-0066-9268}{\protect\includegraphics[height=0.19cm]{figures/orcid.pdf}}}$, 
A.~Bekkelien\inst{\ref{inst:6}}, 
G.~Bruno\inst{\ref{inst:13}}\,$^{\href{https://orcid.org/0000-0002-3288-0802}{\protect\includegraphics[height=0.19cm]{figures/orcid.pdf}}}$, 
H.~P.~Osborn\inst{\ref{inst:8},\ref{inst:14}}\,$^{\href{https://orcid.org/0000-0002-4047-4724}{\protect\includegraphics[height=0.19cm]{figures/orcid.pdf}}}$, 
S.~Ulmer-Moll\inst{\ref{inst:15},\ref{inst:16}}\,$^{\href{https://orcid.org/0000-0003-2417-7006}{\protect\includegraphics[height=0.19cm]{figures/orcid.pdf}}}$, 
T.~G.~Wilson\inst{\ref{inst:17}}\,$^{\href{https://orcid.org/0000-0001-8749-1962}{\protect\includegraphics[height=0.19cm]{figures/orcid.pdf}}}$, 
Y.~Alibert\inst{\ref{inst:8},\ref{inst:7}}\,$^{\href{https://orcid.org/0000-0002-4644-8818}{\protect\includegraphics[height=0.19cm]{figures/orcid.pdf}}}$, 
R.~Alonso\inst{\ref{inst:18},\ref{inst:19}}\,$^{\href{https://orcid.org/0000-0001-8462-8126}{\protect\includegraphics[height=0.19cm]{figures/orcid.pdf}}}$, 
T.~Bárczy\inst{\ref{inst:20}}\,$^{\href{https://orcid.org/0000-0002-7822-4413}{\protect\includegraphics[height=0.19cm]{figures/orcid.pdf}}}$, 
D.~Barrado\inst{\ref{inst:21}}\,$^{\href{https://orcid.org/0000-0002-5971-9242}{\protect\includegraphics[height=0.19cm]{figures/orcid.pdf}}}$, 
S.~C.~C.~Barros\inst{\ref{inst:22},\ref{inst:23}}\,$^{\href{https://orcid.org/0000-0003-2434-3625}{\protect\includegraphics[height=0.19cm]{figures/orcid.pdf}}}$, 
W.~Baumjohann\inst{\ref{inst:12}}\,$^{\href{https://orcid.org/0000-0001-6271-0110}{\protect\includegraphics[height=0.19cm]{figures/orcid.pdf}}}$, 
C.~Broeg\inst{\ref{inst:7},\ref{inst:8}}\,$^{\href{https://orcid.org/0000-0001-5132-2614}{\protect\includegraphics[height=0.19cm]{figures/orcid.pdf}}}$, 
A.~Castro-González\inst{\ref{inst:6}}, 
A.~Collier~Cameron\inst{\ref{inst:24}}\,$^{\href{https://orcid.org/0000-0002-8863-7828}{\protect\includegraphics[height=0.19cm]{figures/orcid.pdf}}}$, 
A.~C.~M.~Correia\inst{\ref{inst:25}}\,$^{\href{https://orcid.org/0000-0002-8946-8579}{\protect\includegraphics[height=0.19cm]{figures/orcid.pdf}}}$, 
Sz.~Csizmadia\inst{\ref{inst:26}}\,$^{\href{https://orcid.org/0000-0001-6803-9698}{\protect\includegraphics[height=0.19cm]{figures/orcid.pdf}}}$, 
P.~E.~Cubillos\inst{\ref{inst:12},\ref{inst:27}}, 
M.~B.~Davies\inst{\ref{inst:28}}\,$^{\href{https://orcid.org/0000-0001-6080-1190}{\protect\includegraphics[height=0.19cm]{figures/orcid.pdf}}}$, 
M.~Deleuil\inst{\ref{inst:29}}\,$^{\href{https://orcid.org/0000-0001-6036-0225}{\protect\includegraphics[height=0.19cm]{figures/orcid.pdf}}}$, 
A.~Deline\inst{\ref{inst:6}}, 
O.~D.~S.~Demangeon\inst{\ref{inst:22},\ref{inst:23}}\,$^{\href{https://orcid.org/0000-0001-7918-0355}{\protect\includegraphics[height=0.19cm]{figures/orcid.pdf}}}$, 
B.-O.~Demory\inst{\ref{inst:8},\ref{inst:30},\ref{inst:7}}\,$^{\href{https://orcid.org/0000-0002-9355-5165}{\protect\includegraphics[height=0.19cm]{figures/orcid.pdf}}}$, 
A.~Derekas\inst{\ref{inst:2}}, 
B.~Edwards\inst{\ref{inst:31}}, 
D.~Ehrenreich\inst{\ref{inst:6},\ref{inst:32}}\,$^{\href{https://orcid.org/0000-0001-9704-5405}{\protect\includegraphics[height=0.19cm]{figures/orcid.pdf}}}$, 
A.~Erikson\inst{\ref{inst:26}}, 
A.~Fortier\inst{\ref{inst:7},\ref{inst:8}}\,$^{\href{https://orcid.org/0000-0001-8450-3374}{\protect\includegraphics[height=0.19cm]{figures/orcid.pdf}}}$, 
L.~Fossati\inst{\ref{inst:12}}\,$^{\href{https://orcid.org/0000-0003-4426-9530}{\protect\includegraphics[height=0.19cm]{figures/orcid.pdf}}}$, 
M.~Fridlund\inst{\ref{inst:33},\ref{inst:34}}\,$^{\href{https://orcid.org/0000-0002-0855-8426}{\protect\includegraphics[height=0.19cm]{figures/orcid.pdf}}}$, 
K.~Gazeas\inst{\ref{inst:35}}\,$^{\href{https://orcid.org/0000-0002-8855-3923}{\protect\includegraphics[height=0.19cm]{figures/orcid.pdf}}}$, 
M.~Gillon\inst{\ref{inst:36}}\,$^{\href{https://orcid.org/0000-0003-1462-7739}{\protect\includegraphics[height=0.19cm]{figures/orcid.pdf}}}$, 
M.~Güdel\inst{\ref{inst:37}}, 
M.~N.~Günther\inst{\ref{inst:38}}\,$^{\href{https://orcid.org/0000-0002-3164-9086}{\protect\includegraphics[height=0.19cm]{figures/orcid.pdf}}}$, 
Ch.~Helling\inst{\ref{inst:12},\ref{inst:39}}, 
K.~G.~Isaak\inst{\ref{inst:38}}\,$^{\href{https://orcid.org/0000-0001-8585-1717}{\protect\includegraphics[height=0.19cm]{figures/orcid.pdf}}}$, 
T.~Keller\inst{\ref{inst:7},\ref{inst:8}}, 
L.~L.~Kiss\inst{\ref{inst:9},\ref{inst:41}}, 
D.~Kitzmann\inst{\ref{inst:7},\ref{inst:8}}, 
J.~Korth\inst{\ref{inst:6}}\,$^{\href{https://orcid.org/0000-0002-0076-6239}{\protect\includegraphics[height=0.19cm]{figures/orcid.pdf}}}$, 
K.~W.~F.~Lam\inst{\ref{inst:26}}\,$^{\href{https://orcid.org/0000-0002-9910-6088}{\protect\includegraphics[height=0.19cm]{figures/orcid.pdf}}}$, 
J.~Laskar\inst{\ref{inst:42}}\,$^{\href{https://orcid.org/0000-0003-2634-789X}{\protect\includegraphics[height=0.19cm]{figures/orcid.pdf}}}$, 
A.~Lecavelier~des~Etangs\inst{\ref{inst:43}}\,$^{\href{https://orcid.org/0000-0002-5637-5253}{\protect\includegraphics[height=0.19cm]{figures/orcid.pdf}}}$, 
A.~Leleu\inst{\ref{inst:6},\ref{inst:7}}\,$^{\href{https://orcid.org/0000-0003-2051-7974}{\protect\includegraphics[height=0.19cm]{figures/orcid.pdf}}}$, 
M.~Lendl\inst{\ref{inst:6}}\,$^{\href{https://orcid.org/0000-0001-9699-1459}{\protect\includegraphics[height=0.19cm]{figures/orcid.pdf}}}$, 
D.~Magrin\inst{\ref{inst:11}}\,$^{\href{https://orcid.org/0000-0003-0312-313X}{\protect\includegraphics[height=0.19cm]{figures/orcid.pdf}}}$, 
P.~F.~L.~Maxted\inst{\ref{inst:44}}\,$^{\href{https://orcid.org/0000-0003-3794-1317}{\protect\includegraphics[height=0.19cm]{figures/orcid.pdf}}}$, 
B.~Merín\inst{\ref{inst:45}}\,$^{\href{https://orcid.org/0000-0002-8555-3012}{\protect\includegraphics[height=0.19cm]{figures/orcid.pdf}}}$, 
C.~Mordasini\inst{\ref{inst:7},\ref{inst:8}}, 
V.~Nascimbeni\inst{\ref{inst:11}}\,$^{\href{https://orcid.org/0000-0001-9770-1214}{\protect\includegraphics[height=0.19cm]{figures/orcid.pdf}}}$, 
D.~Orikhovskyi\inst{\ref{inst:3}}, 
R.~Ottensamer\inst{\ref{inst:37}}, 
I.~Pagano\inst{\ref{inst:13}}\,$^{\href{https://orcid.org/0000-0001-9573-4928}{\protect\includegraphics[height=0.19cm]{figures/orcid.pdf}}}$, 
E.~Pallé\inst{\ref{inst:18},\ref{inst:19}}\,$^{\href{https://orcid.org/0000-0003-0987-1593}{\protect\includegraphics[height=0.19cm]{figures/orcid.pdf}}}$, 
G.~Peter\inst{\ref{inst:26}}\,$^{\href{https://orcid.org/0000-0001-6101-2513}{\protect\includegraphics[height=0.19cm]{figures/orcid.pdf}}}$, 
D.~Piazza\inst{\ref{inst:7}}, 
G.~Piotto\inst{\ref{inst:11},\ref{inst:48}}\,$^{\href{https://orcid.org/0000-0002-9937-6387}{\protect\includegraphics[height=0.19cm]{figures/orcid.pdf}}}$, 
D.~Pollacco\inst{\ref{inst:17}}, 
T.~Pribulla\inst{\ref{inst:3}}, 
D.~Queloz\inst{\ref{inst:14},\ref{inst:49}}\,$^{\href{https://orcid.org/0000-0002-3012-0316}{\protect\includegraphics[height=0.19cm]{figures/orcid.pdf}}}$, 
R.~Ragazzoni\inst{\ref{inst:11},\ref{inst:48}}\,$^{\href{https://orcid.org/0000-0002-7697-5555}{\protect\includegraphics[height=0.19cm]{figures/orcid.pdf}}}$, 
N.~Rando\inst{\ref{inst:38}}, 
H.~Rauer\inst{\ref{inst:50},\ref{inst:51}}\,$^{\href{https://orcid.org/0000-0002-6510-1828}{\protect\includegraphics[height=0.19cm]{figures/orcid.pdf}}}$, 
I.~Ribas\inst{\ref{inst:52},\ref{inst:53}}\,$^{\href{https://orcid.org/0000-0002-6689-0312}{\protect\includegraphics[height=0.19cm]{figures/orcid.pdf}}}$, 
M.~Rieder\inst{\ref{inst:7}}, 
N.~C.~Santos\inst{\ref{inst:22},\ref{inst:23}}\,$^{\href{https://orcid.org/0000-0003-4422-2919}{\protect\includegraphics[height=0.19cm]{figures/orcid.pdf}}}$, 
G.~Scandariato\inst{\ref{inst:13}}\,$^{\href{https://orcid.org/0000-0003-2029-0626}{\protect\includegraphics[height=0.19cm]{figures/orcid.pdf}}}$, 
D.~Ségransan\inst{\ref{inst:6}}\,$^{\href{https://orcid.org/0000-0003-2355-8034}{\protect\includegraphics[height=0.19cm]{figures/orcid.pdf}}}$, 
A.~E.~Simon\inst{\ref{inst:7},\ref{inst:8}}\,$^{\href{https://orcid.org/0000-0001-9773-2600}{\protect\includegraphics[height=0.19cm]{figures/orcid.pdf}}}$, 
A.~M.~S.~Smith\inst{\ref{inst:26}}\,$^{\href{https://orcid.org/0000-0002-2386-4341}{\protect\includegraphics[height=0.19cm]{figures/orcid.pdf}}}$, 
S.~G.~Sousa\inst{\ref{inst:22},\ref{inst:23}}\,$^{\href{https://orcid.org/0000-0001-9047-2965}{\protect\includegraphics[height=0.19cm]{figures/orcid.pdf}}}$, 
R.~Southworth\inst{\ref{inst:38}}, 
M.~Stalport\inst{\ref{inst:16},\ref{inst:36}}, 
M.~Steinberger\inst{\ref{inst:12}}, 
S.~Sulis\inst{\ref{inst:29}}\,$^{\href{https://orcid.org/0000-0001-8783-526X}{\protect\includegraphics[height=0.19cm]{figures/orcid.pdf}}}$, 
S.~Udry\inst{\ref{inst:6}}\,$^{\href{https://orcid.org/0000-0001-7576-6236}{\protect\includegraphics[height=0.19cm]{figures/orcid.pdf}}}$, 
B.~Ulmer\inst{\ref{inst:26}}, 
V.~Van~Grootel\inst{\ref{inst:16}}\,$^{\href{https://orcid.org/0000-0003-2144-4316}{\protect\includegraphics[height=0.19cm]{figures/orcid.pdf}}}$, 
J.~Venturini\inst{\ref{inst:6}}\,$^{\href{https://orcid.org/0000-0001-9527-2903}{\protect\includegraphics[height=0.19cm]{figures/orcid.pdf}}}$, 
E.~Villaver\inst{\ref{inst:18},\ref{inst:19}}, 
V.~Viotto\inst{\ref{inst:11}}, and
N.~A.~Walton\inst{\ref{inst:55}}\,$^{\href{https://orcid.org/0000-0003-3983-8778}{\protect\includegraphics[height=0.19cm]{figures/orcid.pdf}}}$
}

\institute{
\label{inst:2}Eötvös Loránd University, Gothard Astrophysical Observatory, 9700 Szombathely, Szent Imre h. u. 112, Hungary \and
\label{inst:3}Astronomical Institute, Slovak Academy of Sciences, 059 60 Tatranská Lomnica, Slovakia \and
\label{inst:4}Dipartimento di Fisica, Università degli Studi di Torino, via Pietro Giuria 1, 10125 Torino, Italy \and
\label{inst:5}Department of Astronomy, Stockholm University, AlbaNova University Center, 10691 Stockholm, Sweden \and
\label{inst:6}Observatoire astronomique de l'Université de Genève, Chemin Pegasi 51, 1290 Versoix, Switzerland \and
\label{inst:7}Space Research and Planetary Sciences, Physics Institute, University of Bern, Gesellschaftsstrasse 6, 3012 Bern, Switzerland \and
\label{inst:8}Center for Space and Habitability, University of Bern, Gesellschaftsstrasse 6, 3012 Bern, Switzerland \and
\label{inst:9}Konkoly Observatory, HUN-REN Research Centre for Astronomy and Earth Sciences, 1121 Budapest, Konkoly Thege út 15-17, Hungary \and
\label{inst:10}HUN-REN Research Centre for Astronomy and Earth Sciences, MTA Centre of Excellence, 1121 Budapest, Konkoly Thege út 15-17, Hungary \and
\label{inst:11}INAF, Osservatorio Astronomico di Padova, Vicolo dell'Osservatorio 5, 35122 Padova, Italy \and
\label{inst:12}Space Research Institute, Austrian Academy of Sciences, Schmiedlstrasse 6, 8042 Graz, Austria \and
\label{inst:13}INAF, Osservatorio Astrofisico di Catania, Via S. Sofia 78, 95123 Catania, Italy \and
\label{inst:14}ETH Zurich, Department of Physics, Wolfgang-Pauli-Strasse 2, CH-8093 Zurich, Switzerland \and
\label{inst:15}Leiden Observatory, Leiden University, Einsteinweg 55, 2333 CA Leiden, The Netherlands \and
\label{inst:16}Space sciences, Technologies and Astrophysics Research (STAR) Institute, Université de Liège, Allée du 6 Août 19C, 4000 Liège, Belgium \and
\label{inst:17}Department of Physics, University of Warwick, Gibbet Hill Road, Coventry CV4 7AL, United Kingdom \and
\label{inst:18}Instituto de Astrofísica de Canarias, Vía Láctea s/n, 38200 La Laguna, Tenerife, Spain \and
\label{inst:19}Departamento de Astrofísica, Universidad de La Laguna, Astrofísico Francisco Sanchez s/n, 38206 La Laguna, Tenerife, Spain \and
\label{inst:20}Admatis -- Advanced Materials in Space, 5. Kandó Kálmán Street, 3534 Miskolc, Hungary \and
\label{inst:21}Centro de Astrobiología (CSIC-INTA), ESAC campus, 28692 Villanueva de la Cañada (Madrid), Spain \and
\label{inst:22}Instituto de Astrofísica e Ciências do Espaço, Universidade do Porto, CAUP, Rua das Estrelas, 4150-762 Porto, Portugal \and
\label{inst:23}Departamento de Fisica e Astronomia, Faculdade de Ciencias, Universidade do Porto, Rua do Campo Alegre, 4169-007 Porto, Portugal \and
\label{inst:24}Centre for Exoplanet Science, SUPA School of Physics and Astronomy, University of St Andrews, North Haugh, St Andrews KY16 9SS, United Kingdom \and
\label{inst:25}Centro de Física da Universidade de Coimbra, Departamento de Física, Universidade de Coimbra, 3004-516 Coimbra, Portugal \and
\label{inst:26}Institute of Space Research, German Aerospace Center (DLR), Rutherfordstrasse 2, 12489 Berlin, Germany \and
\label{inst:27}INAF, Osservatorio Astrofisico di Torino, Via Osservatorio 20, 10025 Pino Torinese, Italy \and
\label{inst:28}Centre for Mathematical Sciences, Lund University, Box 118, 221 00 Lund, Sweden \and
\label{inst:29}Aix Marseille Univ, CNRS, CNES, LAM, 38 rue Frédéric Joliot-Curie, 13388 Marseille, France \and
\label{inst:30}ARTORG (Artificial Organs) Center for Biomedical Engineering Research, University of Bern, Bern, Switzerland \and
\label{inst:31}Netherlands Institute for Space Research, Niels Bohrweg 4, 2333 CA Leiden, The Netherlands \and
\label{inst:32}Centre Vie dans l’Univers, Faculté des sciences, Université de Genève, Quai Ernest-Ansermet 30, 1211 Genève 4, Switzerland \and
\label{inst:33}Leiden Observatory, Leiden University, PO Box 9513, 2300 RA Leiden, The Netherlands \and
\label{inst:34}Department of Space, Earth and Environment, Chalmers University of Technology, Onsala Space Observatory, 439 92 Onsala, Sweden \and
\label{inst:35}National and Kapodistrian University of Athens, Department of Physics, University Campus, Zografos GR-157 84, Athens, Greece \and
\label{inst:36}Astrobiology Research Unit, Université de Liège, Allée du 6 Août 19C, B-4000 Liège, Belgium \and
\label{inst:37}Department of Astrophysics, University of Vienna, Türkenschanzstrasse 17, 1180 Vienna, Austria \and
\label{inst:38}European Space Agency (ESA), European Space Research and Technology Centre (ESTEC), Keplerlaan 1, 2201 AZ Noordwijk, The Netherlands \and
\label{inst:39}Institute for Theoretical Physics and Computational Physics, Graz University of Technology, Petersgasse 16, 8010 Graz, Austria \and
\label{inst:41}Eötvös Loránd University, Institute of Physics, P\'azm\'any P\'eter s\'et\'any 1/A, 1117 Budapest, Hungary \and
\label{inst:42}Institut de mécanique céleste et de calcul des éphémérides (IMCCE), UMR8028 CNRS, Observatoire de Paris, PSL Université, Sorbonne Université, 77 av. Denfert-Rochereau, 75014 Paris, France \and
\label{inst:43}Institut d'astrophysique de Paris, UMR7095 CNRS, Sorbonne Université, 98bis blvd. Arago, 75014 Paris, France \and
\label{inst:44}Astrophysics Group, Lennard Jones Building, Keele University, Staffordshire, ST5 5BG, United Kingdom \and
\label{inst:45}European Space Agency (ESA), European Space Astronomy Centre (ESAC), Camino Bajo del Castillo s/n, 28692 Villanueva de la Cañada, Madrid, Spain \and
\label{inst:48}Dipartimento di Fisica e Astronomia ``Galileo Galilei'', Università degli Studi di Padova, Vicolo dell'Osservatorio 3, 35122 Padova, Italy \and
\label{inst:49}Cavendish Laboratory, JJ Thomson Avenue, Cambridge CB3 0HE, United Kingdom \and
\label{inst:50}German Aerospace Center (DLR), Markgrafenstrasse 37, 10117 Berlin, Germany \and
\label{inst:51}Institut fuer Geologische Wissenschaften, Freie Universitaet Berlin, Malteserstrasse 74-100, 12249 Berlin, Germany \and
\label{inst:52}Institut de Ciencies de l'Espai (ICE, CSIC), Campus UAB, Can Magrans s/n, 08193 Bellaterra, Spain \and
\label{inst:53}Institut d'Estudis Espacials de Catalunya (IEEC), 08860 Castelldefels, Barcelona, Spain \and
\label{inst:55}Institute of Astronomy, University of Cambridge, Madingley Road, Cambridge, CB3 0HA, United Kingdom
}

\date{Received January 21, 2026; accepted July 25, 2026}

\abstract
{We present new CHEOPS transit observations of AU\,Mic\,b and AU\,Mic\,c obtained between June and September 2024, extending the baseline of transit-timing measurements of this young planetary system. For AU\,Mic\,b, the timing signal is well established, with a semi-amplitude ($10 \pm 3\,\mathrm{min}$) and a characteristic modulation timescale ($1168 \pm 20\,\mathrm{d}$) consistent with previous determinations. By contrast, the new CHEOPS data show that the large transit-timing deviation of AU\,Mic\,c reported previously was not sustained. After the steadily increasing timing trend observed in 2022 and 2023, the 2024 timings returned closer to the zero point of the observed-minus-calculated diagram, indicating a reversal of the previously reported behavior. For AU\,Mic\,c, both the transit-timing semi-amplitude ($46 \pm 26\,\mathrm{min}$) and the characteristic modulation timescale ($2150 \pm 110\,\mathrm{d}$) remain tentative. These results highlight the importance of continued long-term monitoring of the AU\,Mic system.}


\titlerunning{2024 CHEOPS reveals a TTV reversal in AU\,Mic\,c}
\authorrunning{Z. Garai et al.}
\maketitle
\nolinenumbers

\section{Introduction}
\label{intro}

AU\,Microscopii is one of the best-studied young \citep[$\sim$20$\,\mathrm{Myr}$ old; see, e.g.,][]{Galindo1} multi-planet systems. The host star is one of the brightest ($V = 8.7\,\mathrm{mag}$) M dwarfs in the sky \citep{Torres1}, located only $\sim$9.71$\,\mathrm{pc}$ from the Sun \citep{Gaia1}, and a member of the Beta\,Pictoris moving group. AU\,Mic forms a wide common-proper-motion system with the AT\,Mic binary \citep{Barrado1}. AU\,Mic is magnetically active with star spots and frequent flaring \citep[see, e.g.,][]{Torres2}. Moreover, it hosts an edge-on debris disk \citep{Kalas1} with fast-moving structures \citep{Boccaletti1}.

Using high-precision photometry from the Transiting Exoplanet Survey Satellite \citep[TESS;][]{Ricker1}, two Neptune-sized close-in planets, AU\,Mic\,b and AU\,Mic\,c, were discovered transiting their host star AU\,Mic \citep{Plavchan1, Martioli1}. Both planets display transit-time variations \citep[TTVs;][]{Szabo1, Szabo2, Wittrock1, Wittrock2, Boldog1}. The peak-to-peak TTV amplitude of AU\,Mic\,b is $24.5\,\mathrm{min}$ \citep{Boldog1}, in agreement with the predictions of \citet{Szabo2}. In 2023, AU\,Mic\,c exhibited a significant timing deviation of $\sim$80\,$\mathrm{min}$, substantially larger than those observed between 2018 and 2022 \citep{Boldog1}. To explain the observed TTVs, \citet{Wittrock2} proposed the existence of an additional non-transiting planet, AU\,Mic\,d, with a period of $\sim$12.73\,$\mathrm{d}$ and a mass of $\sim$1.0\,$\mathrm{M}_\oplus$, orbiting between planets AU\,Mic\,b and AU\,Mic\,c. Recent dynamical simulations support this interpretation \citep{Boldog1}. Based on radial-velocity observations, \citet{Donati1} also reported a non-transiting planet candidate, AU\,Mic\,e, with an orbital period of $\sim$33.4\,$\mathrm{d}$ and a minimum mass of $\sim$35\,$\mathrm{M}_\oplus$, which may also contribute to the observed TTVs.                 

\begin{table*}
\centering
\caption{Log of 2024 CHEOPS photometric observations of AU\,Mic\,b and AU\,Mic\,c.}
\label{cheopsobslog}
\begin{tabular}{cccccccc}
\noalign{\smallskip}
\hline
\hline
\noalign{\smallskip}
Visit  	 & Start date 			 & End date 			 & Duration       & File  							   & Efficiency           & RMS                 & Number\\
No.    	 & [UTC]      			 & [UTC]    			 & [h]            & key   							   & [\%]                 & [ppm]               & of frames\\   
\noalign{\smallskip}
\hline
\noalign{\smallskip}
\multicolumn{8}{c}{AU\,Mic\,b}\\
b1       & 2024-06-09 18:19      & 2024-06-10 09:03      & 14.73          & \texttt{CH\_PR140071\_TG001901}    & 56.0                 & 5960                & 5943\\   
b2       & 2024-06-26 16:13      & 2024-06-27 07:12      & 15.00          & \texttt{CH\_PR140071\_TG002201}    & 59.4                 & 2980                & 6412\\ 
b3       & 2024-07-13 14:03      & 2024-07-14 04:47      & 14.73          & \texttt{CH\_PR140071\_TG002203}    & 66.8                 & 4540                & 7084\\
b4       & 2024-07-30 13:50      & 2024-07-31 04:34      & 14.73          & \texttt{CH\_PR140071\_TG002204}    & 74.4                 & 4660                & 7889\\
b5       & 2024-08-07 23:48      & 2024-08-08 15:17      & 15.49          & \texttt{CH\_PR140071\_TG002205}    & 81.8                 & 3680                & 9128\\
b6       & 2024-08-16 11:28      & 2024-08-17 01:57      & 14.50          & \texttt{CH\_PR140071\_TG002206}    & 81.5                 & 5520                & 8505\\
b7       & 2024-08-24 21:39      & 2024-08-26 14:21      & 40.69          & \texttt{CH\_PR140071\_TG002207}    & 73.1                 & 10100               & 21413\\
b8       & 2024-09-02 08:46      & 2024-09-03 06:23      & 21.63          & \texttt{CH\_PR140071\_TG002208}    & 68.5                 & 9910                & 10668\\
b9       & 2024-09-10 19:53      & 2024-09-11 10:51      & 14.97          & \texttt{CH\_PR140071\_TG002209}    & 64.3                 & 10760               & 6930\\
\noalign{\smallskip}
\hline
\noalign{\smallskip}
\multicolumn{8}{c}{AU\,Mic\,c}\\
c1       & 2024-06-11 11:10      & 2024-06-12 05:32      & 18.37          & \texttt{CH\_PR140071\_TG001801}    & 58.0                 & 3330                & 7672\\
c2       & 2024-06-30 05:59      & 2024-07-01 00:01      & 18.03          & \texttt{CH\_PR140071\_TG002301}    & 57.1                 & 4300                & 7413\\     
c3       & 2024-07-19 01:37      & 2024-07-19 21:10      & 19.55          & \texttt{CH\_PR140071\_TG002302}    & 69.1                 & 3560                & 9723\\     
c4       & 2024-08-06 22:14      & 2024-08-07 16:51      & 18.62          & \texttt{CH\_PR140071\_TG002303}    & 81.1                 & 5720                & 10871\\     
c5       & 2024-08-24 21:39      & 2024-08-26 14:21      & 40.69          & \texttt{CH\_PR140071\_TG002207}    & 73.1                 & 10100               & 21413\\
\noalign{\smallskip}
\hline
\noalign{\smallskip}
\end{tabular}
\tablefoot{The table gives the time interval of each visit (ISO-8601 format), duration, file key (for retrieval from the CHEOPS archive), efficiency (ratio of effective science time to total visit time), point-to-point root mean square (RMS) of the \texttt{PIPE}-processed light curves, and number of frames.}
\end{table*}

Therefore, further TTV observations are essential to confirm and characterize additional planets in the AU\,Mic system, while updated ephemerides are crucial for planning future follow-up observations \citep[see, e.g.,][]{Yu1}. In particular, continued monitoring of AU\,Mic\,c is important, given its large timing deviation and the uncertain nature of its origin \citep{Boldog1}. In this work, we present photometric observations obtained in 2024 with the Characterizing Exoplanet Satellite \citep[CHEOPS;][]{Benz1,Fortier1} and updated transit-timing-variation results for AU\,Mic\,b and AU\,Mic\,c.        

The paper is organized as follows. In Sect.~\ref{obs}, we describe the CHEOPS observations and data analysis. In Sect.~\ref{res}, we present our results and discuss the updated TTV behavior of AU\,Mic\,b and AU\,Mic\,c. Finally, we summarize our main results in Sect.~\ref{concl}.

\section{CHEOPS observations and data analysis}
\label{obs}

\begin{figure*}
\centering
\centerline{
\includegraphics[width=\columnwidth]{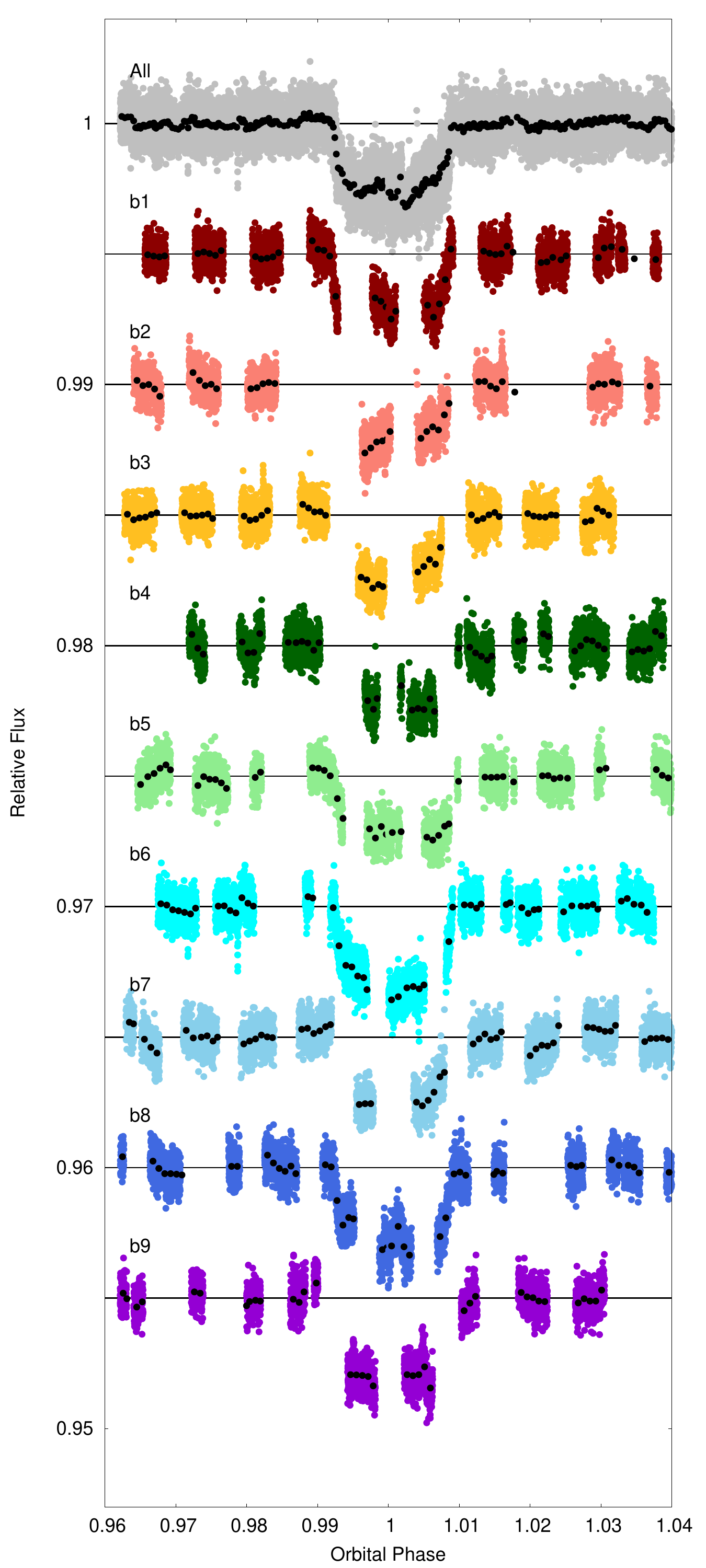}
\includegraphics[width=\columnwidth]{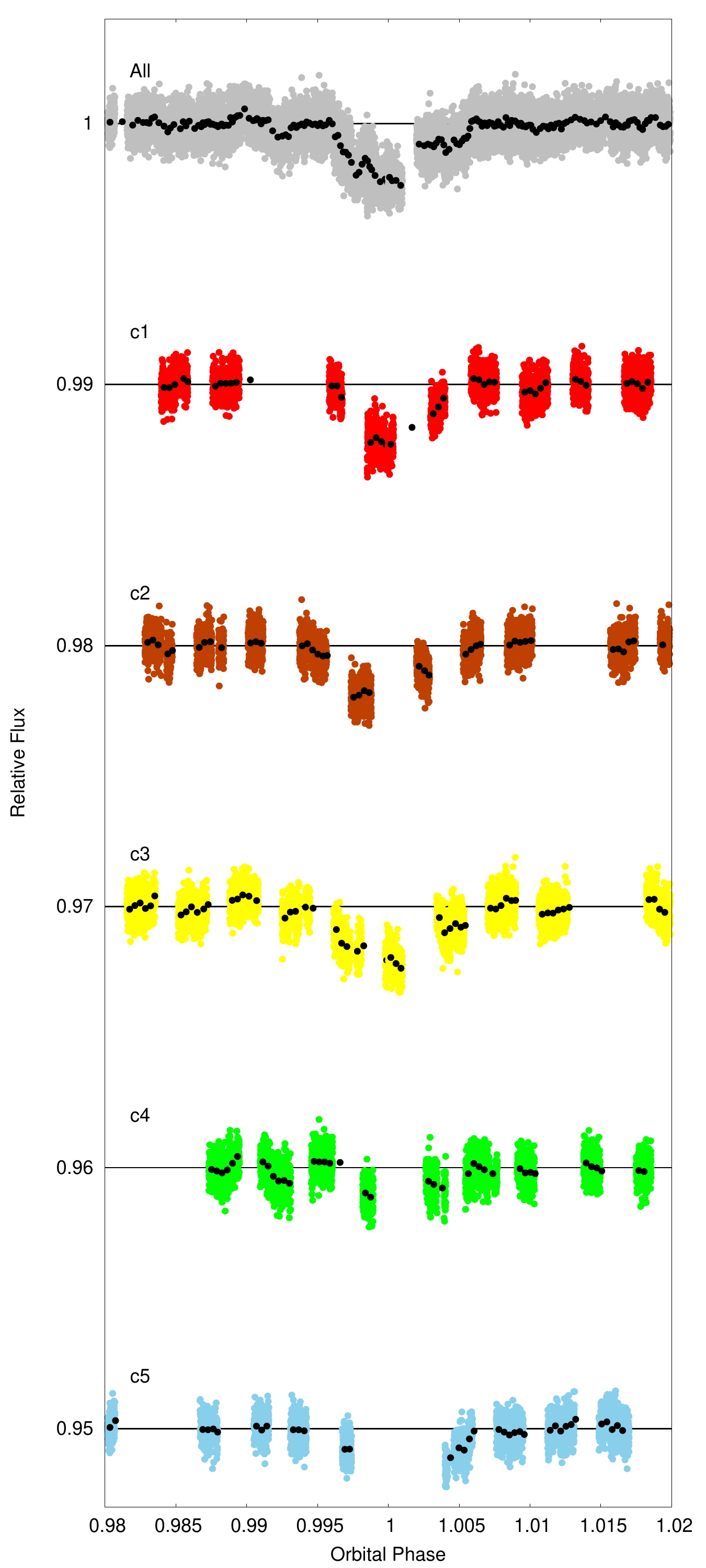}}
\caption{CHEOPS phase-folded and 10-minute-binned transit light curves of AU\,Mic\,b (left) and AU\,Mic\,c (right) from 2024 after detrending with the \texttt{pycheops} software (arbitrarily shifted in relative flux for clarity). The reference mid-transit times and orbital periods used here are $T_\mathrm{c} = 2458330.38416\,\mathrm{BJD_{TDB}}$ and $P_\mathrm{orb} = 8.4631427\,\mathrm{d}$ (planet\,b), and $T_\mathrm{c} = 2459454.8973\,\mathrm{BJD_{TDB}}$ and $P_\mathrm{orb} = 18.85882\,\mathrm{d}$ (planet\,c), based on the results of \citet{Szabo2}. The shape of the combined phase-folded transit profile of AU\,Mic\,c is influenced by the depth variations observed among the individual visits.}
\label{fig:AU_Mic_observations}
\end{figure*}

Between June and September 2024, we used the CHEOPS space telescope to perform nine and five transit observations (visits) of AU\,Mic\,b and AU\,Mic\,c, respectively (see Table\,\ref{cheopsobslog} and Fig.\,\ref{fig:AU_Mic_observations} for more details). In our analysis, the photometry was extracted using the CHEOPS imagettes, which are images centered around the target star with a radius of 30 pixels. The integration time of the imagettes was $5\,\mathrm{s}$. The photometric extraction of the imagettes was carried out using the \texttt{PIPE} (\texttt{PSF Imagette Photometric Extraction}) software, a tool that was specifically developed for this purpose using point-spread function (PSF) photometry \citep{Brandeker1}. The CHEOPS light curves of AU\,Mic were detrended using the dedicated \texttt{pycheops} software \citep{Maxted1}. We first normalized the light curves by unity, removed all flagged points, points with peculiarly high backgrounds (> 20 electrons/pixel), and masked out all flares. The remaining outliers were manually cleaned. Overall, these procedures resulted in the removal or masking of approximately 43.4\% of the data points, primarily due to stellar flares. 

As a next step, long-term trends in the light curves were modeled and removed using a sixth-order polynomial. Because the long-term polynomial detrending of PIPE data in \texttt{pycheops} was performed prior to the transit fit, we independently analyzed the CHEOPS \texttt{DRP} \citep[\texttt{Data Reduction Pipeline}, version 14.1.3;][]{Hoyer1} photometry\footnote{Co-added subarray data products centered around the target star with a radius of 100 pixels \citep{Benz1, Fortier1} and an integration time of $7 \times 5\,\mathrm{s}$. The applied aperture radius was 25 pixels.} using a custom software and fitting procedure in which the polynomial baseline model was fitted simultaneously with the transit model. The resulting transit times were consistent with those obtained from the PIPE photometry and reproduced the same overall TTV behavior for both AU\,Mic\,b and AU\,Mic\,c. This agreement indicates that the reported timing variations and their uncertainties are not driven by the adopted treatment of the long-term baseline. 

We co-fitted the photometric transit models with decorrelation terms based on the extracted parameters, including $x$ and $y$ centroid positions, sine and cosine of roll angle, background flux, time, and similar parameters. We assessed the inclusion of each decorrelation parameter using the Bayes factor method \citep{Swayne1, Maxted1}. Some earlier works \citep[see, e.g.,][]{Osborn1} further detrended the CHEOPS data as a function of roll angle using, for example, a spline function. However, our inspection of the CHEOPS flux residuals as a function of roll angle for each visit revealed no apparent variations that would require such additional modeling. The detrended PIPE dataset was used in the subsequent transit-fitting procedure. 

\begin{table}
\centering
\caption{Priors and best-fitting parameters of AU\,Mic\,b and AU\,Mic\,c from \texttt{Allesfitter}.}
\label{cheops-parameters-tab}
\begin{tabular}{lll}
\noalign{\smallskip}
\hline
\hline
\noalign{\smallskip}
Parameter                                            & Prior                                            & Value\\
\noalign{\smallskip}
\hline
\noalign{\smallskip}
\multicolumn{3}{c}{AU\,Mic\,b}\\
$T_\mathrm{c}$ [$\mathrm{BJD}_\mathrm{TDB}$] 		 & \dots                                            & $2458330.38416^{(a)}$\\
$P_\mathrm{orb}$ [d] 							     & \dots                                            & $8.4631427^{(a)}$\\
$R_\mathrm{p}/R_\mathrm{s}$ 						 & $\mathcal{U}$(0.01, 0.1)       		 	        & $0.04811_{-0.00022}^{+0.00028}$\\
$(R_\mathrm{p} + R_\mathrm{s})/a$ 					 & $\mathcal{U}$(0.05, 0.06)		 		        & $0.0568_{-0.0014}^{+0.0012}$\\
$\cos i$ 										     & $\mathcal{U}$(0.0, 0.0349)		 		        & $0.0088_{-0.0040}^{+0.0056}$\\
$\sqrt{e} \cos \omega$                               & $\mathcal{U}$(-0.2881, 0.2881)                   & $0.04_{-0.15}^{+0.14}$\\
$\sqrt{e} \sin \omega$                               & $\mathcal{U}$(-0.2881, 0.2881)                   & $0.190_{-0.098}^{+0.058}$\\
\noalign{\smallskip}
\hline
\noalign{\smallskip}
\multicolumn{3}{c}{AU\,Mic\,c}\\
$T_\mathrm{c}$ [$\mathrm{BJD}_\mathrm{TDB}$] 		 & \dots                                            & $2459454.8973^{(a)}$\\
$P_\mathrm{orb}$ [d] 							     & \dots                                            & $18.85882^{(a)}$\\
$R_\mathrm{p}/R_\mathrm{s}$ 						 & $\mathcal{U}$(0.01, 0.1)       		 	        & $0.03828_{-0.00050}^{+0.00045}$\\
$(R_\mathrm{p} + R_\mathrm{s})/a$ 					 & $\mathcal{U}$(0.030, 0.036)   		 		    & $0.03507_{-0.00073}^{+0.00059}$\\
$\cos i$ 										     & $\mathcal{U}$(0.0, 0.0349)    		 		    & $0.02715_{-0.00093}^{+0.00082}$\\
$\sqrt{e} \cos \omega$                               & $\mathcal{U}$(-0.2881, 0.2881)                   & $-0.05_{-0.18}^{+0.21}$\\
$\sqrt{e} \sin \omega$                               & $\mathcal{U}$(-0.2881, 0.2881)                   & $-0.237_{-0.038}^{+0.069}$\\
\noalign{\smallskip}
\hline
\noalign{\smallskip}
\multicolumn{3}{c}{LD and GP parameters}\\
$q_1$ 										         & $\mathcal{N}$(0.5100, 0.1)		 			    & $0.486 \pm 0.037$\\
$q_2$					 					         & $\mathcal{N}$(0.2324, 0.1) 					    & $0.139 \pm 0.029$\\
$\ln \sigma$ [$\ln$ rel. flux] 			             & $\mathcal{U}$(-15.0, -5.0) 					    & $-7.6855_{-0.0023}^{+0.0026}$\\ 
$\ln S_\mathrm{0}$		 						     & \dots			                                & $-10.0^{(b)}$\\ 
$\ln Q_\mathrm{0}$		 						     & \dots                 					        & $-0.3465736^{(b)}$\\
$\ln \omega_\mathrm{0}$                 			 & \dots             	 					        & $0.1^{(b)}$\\
\noalign{\smallskip}
\hline
\noalign{\smallskip}
\end{tabular}
\tablefoot{Extra priors: $T_\mathrm{eff} = 3665 \pm 31\,\mathrm{K}$, $R_\mathrm{s} = 0.82 \pm 0.02\,\mathrm{R}_\odot$, $M_\mathrm{s} = 0.60 \pm 0.04\,\mathrm{M}_\odot$ \citep{Donati1}, $\rho_\mathrm{s} = 1.53 \pm 0.16\,\mathrm{g~cm^{-3}}$ (the stellar density prior is automatically set as a normal prior). Parameters without quoted uncertainties were kept fixed during the fitting procedure. $^{(a)}$Based on the results of \citet{Szabo2}. $^{(b)}$Discussed in Sect. \ref{obs}.}
\end{table}

\begin{table}
\centering
\caption{Derived parameters of planets AU\,Mic\,b and AU\,Mic\,c from \texttt{Allesfitter}.}
\label{cheops-parameters-tab2}
\begin{tabular}{ll}
\noalign{\smallskip}
\hline
\hline
\noalign{\smallskip}
Parameter        						& Value\\         
\noalign{\smallskip}
\hline
\noalign{\smallskip}
\multicolumn{2}{c}{AU\,Mic\,b}\\
$R_\mathrm{s}/a$ 						& $0.0542_{-0.0013}^{+0.0012}$\\
$a/R_\mathrm{s}$ 						& $18.46_{-0.39}^{+0.47}$\\ 
$R_\mathrm{p}/a$ 						& $0.002607_{-0.000067}^{+0.000058}$\\ 
$R_\mathrm{p}$ [$\mathrm{R_{\oplus}}$] 	& $4.31 \pm 0.11$\\ 
$R_\mathrm{p}$ [$\mathrm{R_{Jup}}$] 	& $0.3841\pm0.0096$\\ 
$a$ [$\mathrm{R_{\odot}}$] 				& $15.15_{-0.49}^{+0.52}$\\ 
$a$ [au] 								& $0.0705_{-0.0023}^{+0.0024}$\\ 
$i$ [deg] 								& $89.50_{-0.33}^{+0.23}$\\
$e$                                     & $0.058_{-0.027}^{+0.021}$\\
$\omega$ [deg]                          & $84_{-40}^{+52}$\\
$b$ 									& $0.153_{-0.069}^{+0.100}$\\ 
$t_\mathrm{14}^{(a)}$ [h]     	        & $3.4583_{-0.0060}^{+0.0069}$\\ 
$t_\mathrm{23}^{(b)}$ [h]    		    & $3.1314_{-0.012}^{+0.0075}$\\ 
$T_\mathrm{eq}^{(c)}$ [K]		        & $551.5_{-8.1}^{+7.7}$\\ 
$\delta^{(d)}$ [ppm]                    & $2694 \pm 24$\\ 
\noalign{\smallskip}
\hline
\noalign{\smallskip}
\multicolumn{2}{c}{AU\,Mic\,c}\\
$R_\mathrm{s}/a$ 						& $0.03378_{-0.00070}^{+0.00057}$\\
$a/R_\mathrm{s}$ 						& $29.60_{-0.49}^{+0.63}$\\ 
$R_\mathrm{p}/a$ 						& $0.001292_{-0.000032}^{+0.000027}$\\ 
$R_\mathrm{p}$ [$\mathrm{R_{\oplus}}$] 	& $3.421\pm0.095$\\ 
$R_\mathrm{p}$ [$\mathrm{R_{Jup}}$] 	& $0.3052\pm0.0085$\\ 
$a$ [$\mathrm{R_{\odot}}$] 				& $24.31_{-0.72}^{+0.76}$\\ 
$a$ [au] 								& $0.1130_{-0.0034}^{+0.0035}$\\ 
$i$ [deg] 								& $88.444_{-0.048}^{+0.054}$\\
$e$                                     & $0.079_{-0.029}^{+0.036}$\\
$\omega$ [deg]                          & $258_{-32}^{+51}$\\
$b$ 									& $0.8526 \pm 0.0097$\\ 
$t_\mathrm{14}^{(a)}$ [h]			    & $3.067_{-0.032}^{+0.027}$\\ 
$t_\mathrm{23}^{(b)}$ [h]    		    & $2.304_{-0.054}^{+0.046}$\\ 
$T_\mathrm{eq}^{(c)}$ [K] 	            & $435.4_{-5.6}^{+5.2}$\\ 
$\delta^{(d)}$ [ppm]                    & $1353_{-34}^{+31}$\\ 
\noalign{\smallskip}
\hline
\noalign{\smallskip}
\multicolumn{2}{c}{Stellar parameters}\\
$u_\mathrm{1}$ 						    & $0.194_{-0.039}^{+0.035}$\\ 
$u_\mathrm{2}$ 						    & $0.504 \pm 0.055$\\ 
$\rho_\mathrm{s}$ [g~cm$^{-3}$] 		& $1.52_{-0.17}^{+0.20}$\\ 
\noalign{\smallskip}
\hline
\noalign{\smallskip}
\end{tabular}
\tablefoot{$^{(a)}$Total transit duration (first to fourth contact). $^{(b)}$Full transit duration (second to third contact). $^{(c)}$Assuming Bond albedo $A_\mathrm{B} = 0.3$ and uniform heat redistribution. $^{(d)}$Transit depth.}
\end{table}

To derive the orbital and planetary parameters of AU\,Mic\,b and AU\,Mic\,c from the 2024 CHEOPS observations, we employed the \texttt{Allesfitter}\footnote{\url{https://www.allesfitter.com}} software \citep{allesfitter-code, allesfitter-paper} with a nested-sampling algorithm \citep{Speagle1}. We jointly modeled both planets. Several fundamental parameters were optimized during the transit light-curve modeling procedure. A quadratic limb darkening (LD) law was adopted, with $u_1$ and $u_2$ coefficients linearly interpolated from the stellar parameters $T_\mathrm{eff} = 3665 \pm 31\,\mathrm{K}$ and $\log\,g = 4.52 \pm 0.05\,\mathrm{[cgs]}$, reported by \citet{Donati1}, assuming solar metallicity, and using the CHEOPS passband tables of \citet{Claret1}. These LD coefficients were converted to $q_1$ and $q_2$ \citep{Kipping1}, which served as priors to constrain the stellar limb darkening during the fitting procedure. To model the flux baseline, we applied a Gaussian process (GP) regression with the \texttt{SHOTerm} (simple harmonic oscillator) plus \texttt{JitterTerm} kernel, implemented in \texttt{Celerite}\footnote{\url{https://celerite.readthedocs.io/en/stable}} \citep{Foreman1}, fixing the quality factor to $Q_0 = 1/\sqrt{2}$ \citep[see, e.g.,][]{Osborn1}, the frequency to $\ln \omega_0 = 0.1$, and the scaled power to $\ln S_0 = -10$. We tested several GP configurations to assess the impact of correlated noise on the transit-timing information. The adopted GP setup guards against over-fitting of the transits and improves the residuals while leaving the derived mid-transit times unchanged, confirming that our results are robust against reasonable variations in the noise model. The instrumental noise was sampled with the $\ln \sigma$ parameter, and the eccentricity priors were chosen to match the prior of \citet{Eylen1}, which is a zero-mean half-Gaussian with $\sigma = 0.083$ for multiple-transit systems. Using \texttt{Allesfitter}, we quantified the TTVs by allowing the mid-transit time of each 2024 CHEOPS light curve to vary as free parameter, defined with uniform priors: $\mathcal{U}(-0.02, 0.02)\,\mathrm{days}$ for AU\,Mic\,b, and $\mathcal{U}(-0.07, 0.07)\,\mathrm{days}$ for AU\,Mic\,c. During this calculation, the reference mid-transit times $T_\mathrm{c}$ and orbital periods $P_\mathrm{orb}$ were kept fixed to the values found by \citet{Szabo2}. To speed up the computation, the light curves were cut to 12-hour windows centered on each transit. The adopted priors and modeled parameters are listed in Table\,\ref{cheops-parameters-tab}, and the derived parameters in Table\,\ref{cheops-parameters-tab2}. The corresponding transit light curves, overplotted with the inferred \texttt{Allesfitter} model, are shown in Figs.\,\ref{fig:AU_Mic_planetary_transits_joint1} and \ref{fig:AU_Mic_planetary_transits_joint2}. 

For planets AU\,Mic\,b and AU\,Mic\,c, we then calculated the updated linear ephemeris model by performing a least-squares minimisation over the epochs of all available transit events (i.e., not only the 2024 CHEOPS transits; see below). We obtained $T_\mathrm{c} = 2458330.38050 \pm 0.00034\,\mathrm{BJD_{TDB}}$ and $P_\mathrm{orb} = 8.4631745 \pm 0.0000023\,\mathrm{d}$ for planet AU\,Mic\,b, and $T_\mathrm{c} = 2458342.22320 \pm 0.00037\,\mathrm{BJD_{TDB}}$ and $P_\mathrm{orb} = 18.859065 \pm 0.000011\,\mathrm{d}$ for planet AU\,Mic\,c. The longer time baseline compared to the previous CHEOPS-based results \citep{Szabo1, Szabo2, Boldog1} improves the robustness of the period estimates, particularly for AU\,Mic\,c, where a single CHEOPS season covers only a few transits. To quantify the TTVs, we finally derived the observed-minus-calculated ($O-C$) values by subtracting the linear ephemeris model from the single transit times. The $O-C$ values, including literature values, are listed in Tables\,\ref{table:ttvsb} and \ref{table:ttvsc}, and visualized in Fig. \ref{fig:AU_Mic_ttv}.     

\begin{table}
\centering
\caption{Observed mid-transit times and $O-C$ values of AU\,Mic\,b from TESS, \textit{Spitzer}, and CHEOPS.}
\label{table:ttvsb}
\begin{tabular}{lcr}
\noalign{\smallskip}
\hline
\hline
\noalign{\smallskip}
Observation                         & Transit time                      & $O-C$\\
                                    & [$\mathrm{BJD_{TDB}}-2450000$]    & [min]\\
\noalign{\smallskip}
\hline
\noalign{\smallskip}
TESS S1\#{}1$^{(a)}$                & 8330.39111$\pm$0.00090            & $15.3\pm1.3$\\
TESS S1\#{}2$^{(a)}$                & 8347.31740$\pm$0.00090            & $15.2\pm1.3$\\
\textit{Spitzer}\#{}1$^{(b)}$       & 8525.0451$\pm$0.0010              & $16.7\pm1.4$\\
TESS S27\#{}1$^{(a)}$               & 9041.28162$\pm$0.00083            & $-8.0\pm1.2$\\
TESS S27\#{}2$^{(a)}$               & 9049.74570$\pm$0.00083            & $-6.7\pm1.2$\\
TESS S27\#{}3$^{(a)}$               & 9058.20803$\pm$0.00083            & $-7.9\pm1.2$\\
CHEOPS 20-07-10$^{(a)}$             & 9041.28283$\pm$0.00060            & $-6.3\pm0.87$\\
CHEOPS 20-08-21$^{(a)}$             & 9083.59701$\pm$0.00040            & $-8.7\pm0.58$\\
CHEOPS 20-09-24$^{(a)}$             & 9117.45150$\pm$0.00083            & $-6.1\pm 1.2$\\
CHEOPS 21-07-26$^{(c)}$             & 9422.1342$\pm$0.0010              & $6.0\pm1.4$\\
CHEOPS 21-08-12$^{(c)}$             & 9439.0636$\pm$0.0021              & $10.4\pm3.2$\\
CHEOPS 21-08-29$^{(c)}$             & 9455.98951$\pm$0.00070            & $9.8\pm1.0$\\
CHEOPS 21-09-06$^{(c)}$             & 9464.45311$\pm$0.00090            & $10.4\pm1.3$\\
CHEOPS 22-08-08$^{(d)}$             & 9769.1264$\pm$0.0013              & $9.0\pm1.9$\\
CHEOPS 22-08-25$^{(d)}$             & 9786.0532$\pm$0.0013              & $9.6\pm1.9$\\
CHEOPS 22-08-28$^{(d)}$             & 9819.9045$\pm$0.0013              & $7.6\pm1.9$\\
CHEOPS 23-08-10$^{(d)}$             & 10166.8751$\pm$0.0037             & $-20.6\pm 5.2$\\
CHEOPS 23-08-27$^{(d)}$             & 10183.8089$\pm$0.0055             & $-9.8\pm7.9$\\
CHEOPS 23-09-13$^{(d)}$             & 10200.7339$\pm$0.0064             & $-11.8\pm9.2$\\
CHEOPS 24-06-09$^{(e)}$             & 10471.56232$\pm$0.00083           & $-1.9\pm1.2$\\ 
CHEOPS 24-06-26$^{(e)}$             & 10488.4878$\pm$0.0017             & $-3.1\pm2.4$\\ 
CHEOPS 24-07-13$^{(e)}$             & 10505.4175$\pm$0.0056             & $1.7\pm8.1$\\
CHEOPS 24-07-30$^{(e)}$             & 10522.3458$\pm$0.0044             & $4.5\pm6.3$\\
CHEOPS 24-08-07$^{(e)}$             & 10530.8078$\pm$0.0015             & $2.9\pm2.2$\\
CHEOPS 24-08-16$^{(e)}$             & 10539.27118$\pm$0.00076           & $3.1\pm1.1$\\ 
CHEOPS 24-08-24$^{(e)}$             & 10547.7365$\pm$0.0040             & $6.2\pm 5.8$\\ 
CHEOPS 24-09-02$^{(e)}$             & 10556.19736$\pm$0.00076           & $2.8\pm 1.1$\\ 
CHEOPS 24-09-10$^{(e)}$             & 10564.6616$\pm$0.0068             & $4.4\pm 9.8$\\
\noalign{\smallskip}
\hline
\noalign{\smallskip}
\end{tabular}
\tablefoot{$O-C$ values were computed using $T_c = 2458330.38050\,\mathrm{BJD_{TDB}}$ and $P_\mathrm{orb} = 8.4631745\,\mathrm{d}$. References: $^{(a)}$\citet{Szabo1}, $^{(b)}$\citet{Plavchan1}, $^{(c)}$\citet{Szabo2}, $^{(d)}$\citet{Boldog1}, $^{(e)}$This work.}
\end{table}

\begin{table}
\centering
\caption{Observed mid-transit times and $O-C$ values of AU\,Mic\,c from TESS and CHEOPS.}
\label{table:ttvsc}
\begin{tabular}{lcr}
\noalign{\smallskip}
\hline
\hline
\noalign{\smallskip}
Observation                         & Transit time                      & $O-C$\\
                                    & [$\mathrm{BJD_{TDB}}-2450000$]    & [min]\\
\noalign{\smallskip}
\hline
\noalign{\smallskip}
TESS S1\#{}1$^{(a)}$                & 8342.22432$\pm$0.00050            & $1.6\pm 0.71$\\
TESS S27\#{}1$^{(a)}$               & 9040.00697$\pm$0.00061            & $-2.4\pm0.88$\\
TESS S27\#{}2$^{(a)}$               & 9058.86596$\pm$0.00068            & $-2.5\pm1.0$\\
CHEOPS 21-08-09$^{(b)}$             & 9436.0323$\pm$0.0045              & $-24.0\pm 6.5$\\
CHEOPS 21-08-28$^{(b)}$             & 9454.8988$\pm$0.0050              & $-13.3\pm5.5$\\
CHEOPS 22-08-02$^{(c)}$             & 9794.3646$\pm$0.0032              & $-9.6\pm4.7$\\
CHEOPS 22-08-21$^{(c)}$             & 9813.2383$\pm$0.0031              & $11.5\pm4.6$\\
CHEOPS 23-07-08$^{(c)}$             & 10133.8682$\pm$0.0028             & $48.6\pm4.0$\\
CHEOPS 23-07-27$^{(c)}$             & 10152.7331$\pm$0.0036             & $57.0\pm5.2$\\
CHEOPS 24-06-11$^{(d)}$             & 10473.2893$\pm$0.0073             & $-11.8\pm10.5$\\  
CHEOPS 24-06-30$^{(d)}$             & 10492.1319$\pm$0.0047             & $-35.7\pm6.8$\\
CHEOPS 24-07-19$^{(d)}$             & 10511.0105$\pm$0.0073             & $-7.5\pm10.5$\\
CHEOPS 24-08-06$^{(d)}$             & 10529.8599$\pm$0.0086             & $-21.4\pm12.4$\\ 
CHEOPS 24-08-24$^{(d)}$             & 10548.7073$\pm$0.0088             & $-38.2\pm12.7$\\
\noalign{\smallskip}
\hline
\noalign{\smallskip}
\end{tabular}
\tablefoot{$O-C$ values were computed using $T_c = 2458342.22320\,\mathrm{BJD_{TDB}}$ and $P_\mathrm{orb} = 18.859065\,\mathrm{d}$. References: $^{(a)}$\citet{Gilbert1}, $^{(b)}$\citet{Szabo2}, $^{(c)}$\citet{Boldog1}, $^{(d)}$This work.}
\end{table}

\begin{figure*}
\centering
\centerline{
\includegraphics[width=\columnwidth]{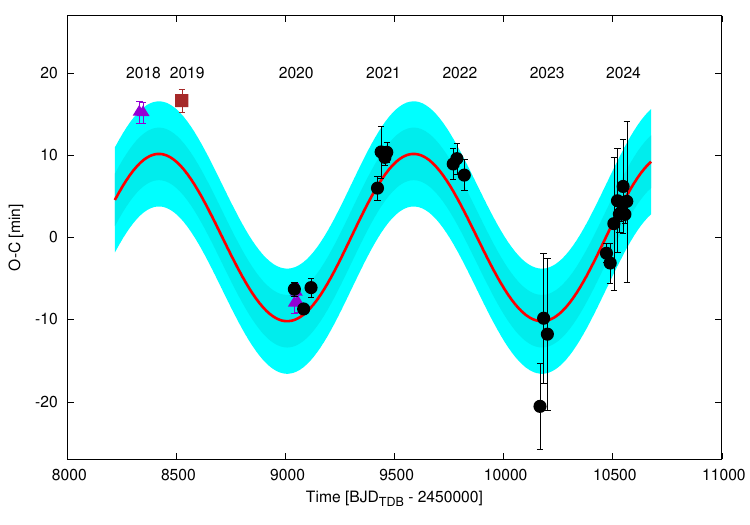}
\includegraphics[width=\columnwidth]{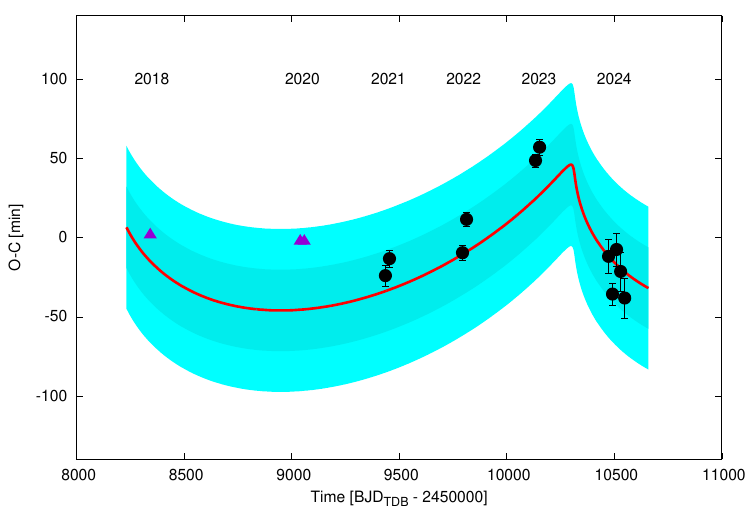}}
\caption{Observed-minus-calculated ($O-C$) diagrams of mid-transit times for AU\,Mic\,b (left panel) and AU\,Mic\,c (right panel), computed using the updated linear ephemerides from this work $T_\mathrm{c} = 2458330.38050 \pm 0.00034\,\mathrm{BJD_{TDB}}$, $P_\mathrm{orb} = 8.4631745 \pm 0.0000023\,\mathrm{d}$ (planet\,b), and $T_\mathrm{c} = 2458342.22320 \pm 0.00037\,\mathrm{BJD_{TDB}}$, $P_\mathrm{orb} = 18.859065 \pm 0.000011\,\mathrm{d}$ (planet\,c), with the best-fitting \texttt{OCFit} models, indicated with red lines. The $1\sigma$ and $2\sigma$ uncertainties of the models are plotted as colored areas in cyan. TESS, \textit{Spitzer}, and CHEOPS observations are depicted with violet triangles, brown square, and black circles, respectively. All data points are plotted with $y$-errorbars. Calendar years are also indicated for easier orientation in the dataset.}
\label{fig:AU_Mic_O-C_fit}
\end{figure*}

The entire $O-C$ dataset (i.e., all observed mid-transit times) was further analyzed with the \texttt{OCFit} code, version 0.2.1\footnote{\url{https://github.com/pavolgaj/OCFit}} \citep{Gajdos1, Gajdos2}. We modeled the $O-C$ data with the \texttt{LiTE3} package, which assumes a light-time effect caused by a third body \citep{Irwin1}. In the present phenomenological application, the fitted period of the \texttt{LiTE3} model is used only as a characteristic timescale of the long-term modulation in the $O-C$ diagram. We do not interpret this parameter as the orbital period of an actual third body or as the resonant TTV superperiod of the system. The initial parameter values were estimated with Genetic algorithms \citep[][]{Hartmann1} and refined using a Markov chain Monte Carlo approach. The final parameter estimates were obtained after $10^6$ iterations. The \texttt{OCFit} best-fitting parameters are given in Table\,\ref{ocfit-parameters-tab}, and the $O-C$ diagrams with best-fitting models are shown in Fig.\,\ref{fig:AU_Mic_O-C_fit}.             

\section{Results and discussion}
\label{res}

The orbital and planetary parameters of AU\,Mic\,b and AU\,Mic\,c are summarized in Tables\,\ref{cheops-parameters-tab} and \ref{cheops-parameters-tab2}, while a comparison with literature values is given in Tables~\ref{tab:solutionsb} and \ref{tab:solutionsc}. Most parameters are consistent with previous determinations within the quoted uncertainties. Here we briefly focus on the planet-to-star radius ratio, $R_\mathrm{p}/R_\mathrm{s}$. Based on the 2024 CHEOPS data, we find that the $R_\mathrm{p}/R_\mathrm{s}$ value of AU\,Mic\,b is significantly ($> 3\sigma$) smaller than the value measured in 2023. In contrast, AU\,Mic\,c shows consistent ($< 3\sigma$) $R_\mathrm{p}/R_\mathrm{s}$ values in 2024 and 2023. These differences may reflect the influence of stellar activity, particularly the evolving spot distribution on AU\,Mic and its impact on the observed transit shapes \citep{Szabo1}.

The apparent difference between the depth variations of AU\,Mic\,b and AU\,Mic\,c should be interpreted with caution. The transits of AU\,Mic\,c are shallower and fewer in number, resulting in larger uncertainties in $R_\mathrm{p}/R_\mathrm{s}$ than for AU\,Mic\,b. Therefore, depth variations comparable to those observed for AU\,Mic\,b could remain hidden within the current confidence intervals of AU\,Mic\,c. Nevertheless, the different impact parameters of the two planets imply that they probe different stellar chords. If the observed depth variations are primarily driven by occulted active regions rather than by disk-integrated spot coverage, the stronger apparent variability of AU\,Mic\,b may indicate that its transit chord crossed a more heterogeneous distribution of stellar active regions during the observed epochs. A detailed investigation of this possibility is beyond the scope of the present work, but simultaneous monitoring of both planets may provide valuable constraints on the spatial and temporal evolution of stellar activity on AU\,Mic. We also examined the relation between the raw stellar flux level and the individual transit depths measured during the 2024 campaign. Neither planet shows a statistically significant correlation between the stellar flux level and the measured transit depth, likely owing to the limited number of observations and the evolving spot distribution on the stellar surface.

Our analysis confirms the presence of TTVs in both AU\,Mic\,b and AU\,Mic\,c, with a higher peak-to-peak amplitude and a non-sinusoidal pattern for AU\,Mic\,c compared to AU\,Mic\,b (see Fig.\,\ref{fig:AU_Mic_O-C_fit}). Using the \texttt{OCFit} code, we derived a semi-amplitude of $A_\mathrm{TTV,b} = 10 \pm 3\,\mathrm{min}$ for AU\,Mic\,b and a tentative semi-amplitude of $A_\mathrm{TTV,c} = 46 \pm 26\,\mathrm{min}$ for AU\,Mic\,c. The result for AU\,Mic\,b is broadly consistent with previous determinations \citep{Szabo2, Boldog1}. The \texttt{LiTE3} models yield characteristic modulation timescales of $P_\mathrm{mod,b}=1168\pm20\,\mathrm{d}$ and $P_\mathrm{mod,c}=2150\pm110\,\mathrm{d}$ for AU\,Mic\,b and AU\,Mic\,c, respectively. In the case of AU\,Mic\,b, the recovered modulation timescale is similar to the 1150-day timescale reported by \citet{Boldog1}. For AU\,Mic\,c, the inferred modulation timescale remains highly tentative given the limited temporal coverage, the irregular nature of the observed TTVs, and the large eccentricity implied by the fit (see Table \ref{ocfit-parameters-tab}). The large eccentricity returned by the \texttt{LiTE3} fit ($e \approx 0.93$) should not be interpreted as evidence for a highly eccentric physical perturber. Rather, it reflects the flexibility required by the adopted phenomenological model to reproduce the irregular shape of the observed $O-C$ curve. The model should therefore be regarded primarily as a phenomenological description of the long-term trend in the $O-C$ diagram rather than as a physical dynamical solution. A comprehensive dynamical analysis of the AU\,Mic system is beyond the scope of the present work. The primary objective of this paper is to present the 2024 CHEOPS timing measurements and assess their impact on the previously reported TTV behavior. Future dynamical modeling incorporating the expanding 2025 and 2026 CHEOPS datasets, together with newly available TESS observations, will be required to place stronger constraints on the architecture of the system and to evaluate the compatibility of the TTV signals observed for AU\,Mic\,b and AU\,Mic\,c.

The new 2024 CHEOPS data reveal notable changes in the temporal evolution of the TTVs for AU\,Mic\,c. While the 2022 and 2023 CHEOPS data suggested a steadily increasing $O-C$ trend \citep{Boldog1}, with maximum deviations of up to $\sim$60\,$\mathrm{min}$ in 2023 (according to the updated linear ephemeris, see Sect. \ref{obs}), the 2024 transits generally clustered closer to the zero point of the $O-C$ diagram, as shown in Fig.\,\ref{fig:AU_Mic_O-C_fit} (right)\footnote{The seasonal RMS scatter of the AU\,Mic\,c timing measurements is approximately $6-15\,\mathrm{min}$, substantially smaller than the $\sim$76-$\mathrm{minute}$ difference between the mean $O-C$ levels observed in 2023 and 2024. This comparison suggests that the large 2023 deviation cannot be explained solely by the typical intra-season scatter of the timing measurements.}. In 2024, the $O-C$ values of planet AU\,Mic\,c span from $-38 \pm 13\,\mathrm{min}$ to $-8 \pm 11\,\mathrm{min}$. This apparent change in trend could indicate that AU\,Mic\,c’s timing signal is subject to additional dynamical perturbations \citep{Wittrock2, Donati1} not captured by the simple \texttt{LiTE3} model. Although stellar activity can affect transit-timing measurements in AU\,Mic \citep{Szabo1}, the spot-induced timing variations reported for AU\,Mic\,b are typically only a few minutes and therefore cannot easily account for the full amplitude of the 2023 AU\,Mic\,c deviation. Consequently, the large offset observed in 2023 may reflect a combination of dynamical perturbations, stellar activity, light-curve morphology, incomplete transit coverage, or residual systematics in the individual visits. A dedicated reanalysis of the 2023 light curves is beyond the scope of the present work, but the behavior observed in 2024 demonstrates that the large 2023 deviation was not sustained and highlights the importance of continued long-term monitoring. If the 2023 measurements are interpreted as outliers, the actual TTV amplitude of AU\,Mic\,c may be substantially smaller than previously suggested and potentially approach the few-minute level predicted by \citet{Szabo2}.

\section{Conclusions}
\label{concl}

The 2024 CHEOPS campaign extends the baseline of AU\,Mic transit-timing measurements and strengthens the evidence for TTVs in this system. The timing signal of AU\,Mic\,b is well established and consistent with previous determinations, reinforcing the picture of a stable and coherent modulation. By contrast, the behavior of AU\,Mic\,c remains uncertain: while the 2022--2023 data suggested a steadily increasing $O-C$ trend with large timing deviation in 2023, the 2024 observations show a return toward the zero point of the $O-C$ diagram, indicating that the large 2023 deviation was not sustained in 2024. Although we did not perform dynamical modeling here, the origin of the large 2023 deviation remains uncertain. Possible explanations include additional dynamical perturbations, stellar activity acting in combination with observational or modeling effects, or a combination of these factors. As emphasized in earlier works \citep{Szabo1, Szabo2, Boldog1}, sustained long-term monitoring with CHEOPS and other high-precision facilities, combined with future dynamical modeling, will be essential to clarify the true nature of the observed variations and to constrain the architecture of the AU\,Mic planetary system.

\section*{Data availability}

The AU\,Mic photometry data used in this work are available in electronic form at the CDS through anonymous ftp to \url{cdsarc.u-strasbg.fr} (130.79.128.5) or by \url{http://cdsweb.u-strasbg.fr/cgi-bin/qcat?J/A+A/}.

\begin{acknowledgements}
The authors thank José A. Caballero, as well as the anonymous reviewer, for constructive comments and suggestions. CHEOPS is an ESA mission in partnership with Switzerland with important contributions to the payload and the ground segment from Austria, Belgium, France, Germany, Hungary, Italy, Portugal, Spain, Sweden, and the United Kingdom. The CHEOPS Consortium would like to gratefully acknowledge the support received by all the agencies, offices, universities, and industries involved. Their flexibility and willingness to explore new approaches were essential to the success of this mission. CHEOPS data analyzed in this article will be made available in the CHEOPS mission archive (\url{https://cheops.unige.ch/archive_browser/}). ZG and GyMSz acknowledge the support from the ESA PRODEX projects Nos. 4000137122, 4000149203, and 4000149202, as well as from the VEGA grant No. 2/0033/26, the contract No. APVV-24-0160, the support from SNN-147362 and the ADVANCED-153410 of the National Research, Development and Innovation Office (NKFIH, Hungary), and the support of the city of Szombathely. This work was supported by the bilateral mobility project No. NKM2024-37/HAS-SAS-2024-3. DG gratefully acknowledges financial support from the CRT foundation under Grant No. 2018.2323. ABr was supported by the SNSA. LBo, GBr, VNa, IPa, GPi, RRa, and GSc acknowledge support from CHEOPS ASI-INAF agreement No. 2019-29-HH.0. This work has been carried out within the framework of the NCCR PlanetS supported by the Swiss National Science Foundation under grants 51NF40\_182901 and 51NF40\_205606. TWi acknowledges support from the UKSA and the University of Warwick. YAl acknowledges support from the Swiss National Science Foundation (SNSF) under grant 200020\_192038. DB, EP, EV, IR and RA acknowledge financial support from the Agencia Estatal de Investigación of the Ministerio de Ciencia e Innovación MCIN/AEI/10.13039/501100011033 and the ERDF “A way of making Europe” through projects PID2021-125627OB-C31, PID2021-125627OB-C32, PID2021-127289NB-I00, PID2023-150468NB-I00 and PID2023-149439NB-C41 from the Centre of Excellence “Severo Ochoa'' award to the Instituto de Astrofísica de Canarias (CEX2019-000920-S), the Centre of Excellence “María de Maeztu” award to the Institut de Ciències de l’Espai (CEX2020-001058-M), and from the Generalitat de Catalunya/CERCA programme. SCCB acknowledges the support from Fundação para a Ciência e Tecnologia (FCT) in the form of work contract through the Scientific Employment Incentive program with reference 2023.06687.CEECIND and DOI 10.54499/2023.06687.CEECIND/CP2839/CT0002. C.B. acknowledges support from the Swiss Space Office through the ESA PRODEX program. ACC acknowledges support from STFC consolidated grant number ST/V000861/1 and UKRI/ERC Synergy Grant EP/Z000181/1 (REVEAL). ACMC acknowledges support from the FCT, Portugal, through the CFisUC projects UIDB/04564/2020 and UIDP/04564/2020, with DOI identifiers 10.54499/UIDB/04564/2020 and 10.54499/UIDP/04564/2020, respectively. A.C., A.D., B.E., K.G., and J.K. acknowledge their role as ESA-appointed CHEOPS Science Team Members. P.E.C. is funded by the Austrian Science Fund (FWF) Erwin Schroedinger Fellowship, program J4595-N. This project was supported by the CNES. This work was supported by FCT - Funda\c{c}\~{a}o para a Ci\^{e}nciae a Tecnologia through national funds and by FEDER through COMPETE2020 through the research grants UIDB/04434/2020, UIDP/04434/2020, 2022.06962.PTDC. O.D.S.D. is supported in the form of work contract (DL 57/2016/CP1364/CT0004) funded by national funds through FCT. B.-O.D. acknowledges support from the Swiss State Secretariat for Education, Research and Innovation (SERI) under contract number MB22.00046. This project has received funding from the Swiss National Science Foundation for project 200021\_200726. MF and CMP gratefully acknowledge the support of the Swedish National Space Agency (DNR 65/19, 174/18). MG is F.R.S.-FNRS Research Director. MNG is the ESA CHEOPS Project Scientist and Mission Representative. BMM is the ESA CHEOPS Project Scientist. KGI was the ESA CHEOPS Project Scientist until the end of December 2022 and Mission Representative until the end of January 2023. All of them are/were responsible for the Guest Observers (GO) Programme. None of them relay/relayed proprietary information between the GO and Guaranteed Time Observation (GTO) Programmes, nor do/did they decide on the definition and target selection of the GTO Programme. Calculations were performed using supercomputer resources provided by the Vienna Scientific Cluster (VSC). J.K. acknowledges support from the Swiss National Science Foundation under grant number TMSGI2\_211697. K.W.F.L. was supported by Deutsche Forschungsgemeinschaft grants RA714/14-1 within the DFG Schwerpunkt SPP 1992, Exploring the Diversity of Extrasolar Planets. This work was granted access to the HPC resources of MesoPSL financed by the Region Ile de France and the project Equip@Meso (reference ANR-10-EQPX-29-01) of the programme Investissements d'Avenir supervised by the Agence Nationale pour la Recherche. AL acknowledges support of the Swiss National Science Foundation under grant number  TMSGI2\_211697. ML acknowledges support of the Swiss National Science Foundation under grant number PCEFP2\_194576. PM acknowledges support from STFC research grant number ST/R000638/1. This work was also partially supported by a grant from the Simons Foundation (PI Queloz, grant number 327127). NCSa acknowledges funding by the European Union (ERC, FIERCE, 101052347). Views and opinions expressed are however those of the author(s) only and do not necessarily reflect those of the European Union or the European Research Council. Neither the European Union nor the granting authority can be held responsible for them. A.S. acknowledges support from the Swiss Space Office through the ESA PRODEX program. S.G.S. acknowledge support from FCT through FCT contract nr. CEECIND/00826/2018 and POPH/FSE (EC). The Portuguese team thanks the Portuguese Space Agency for the provision of financial support in the framework of the PRODEX Programme of the European Space Agency (ESA) under contract number 4000142255. V.V.G. is an F.R.S-FNRS Research Associate. JV acknowledges support from the Swiss National Science Foundation (SNSF) under grant PZ00P2\_208945. NAW acknowledges UKSA grant ST/R004838/1.
\end{acknowledgements}

\bibliographystyle{aa} 
\bibliography{Yourfile}

\begin{appendix}

\section{Additional figures}
\label{app:figures}

This appendix contains additional figures supporting the analyses presented in the main text. 

\begin{@twocolumnfalse}

\begin{figure}[h]
\parbox{\textwidth}{
\centering
\centerline{
\includegraphics[width=\textwidth/3]{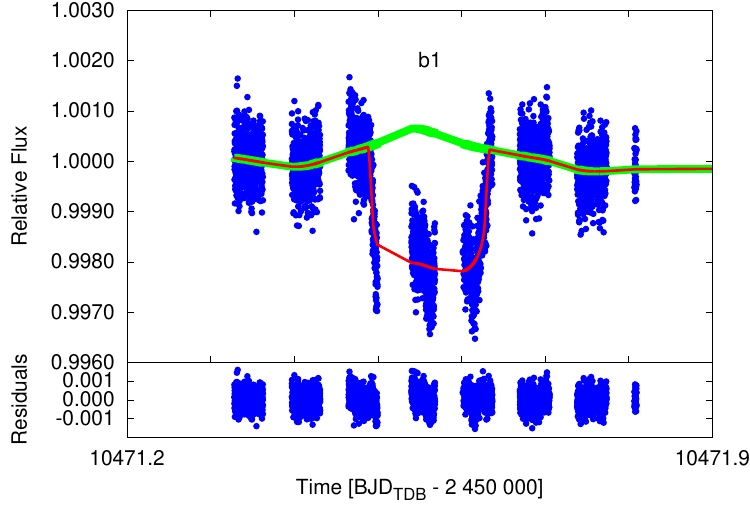}
\includegraphics[width=\textwidth/3]{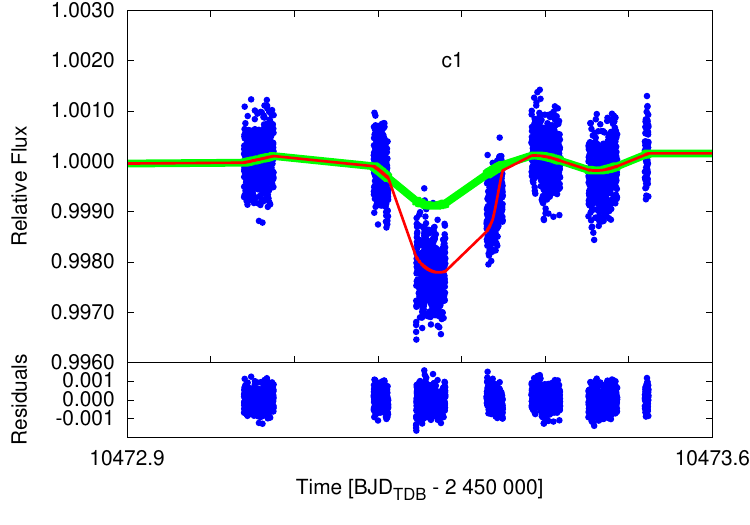}
\includegraphics[width=\textwidth/3]{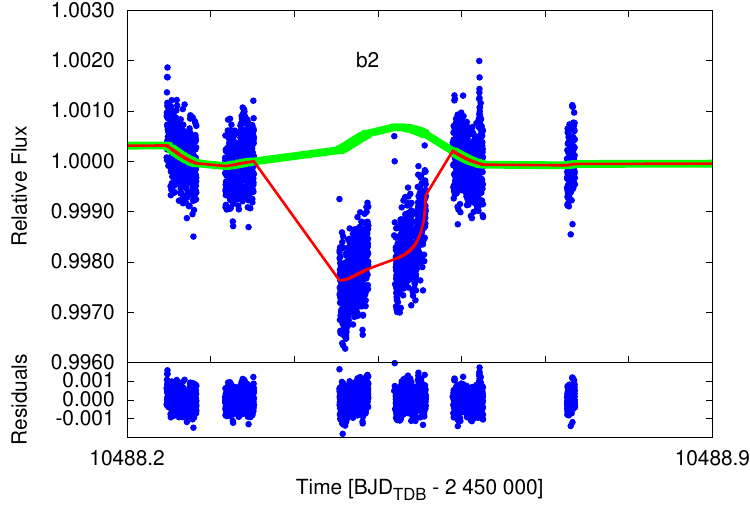}}
\centerline{
\includegraphics[width=\textwidth/3]{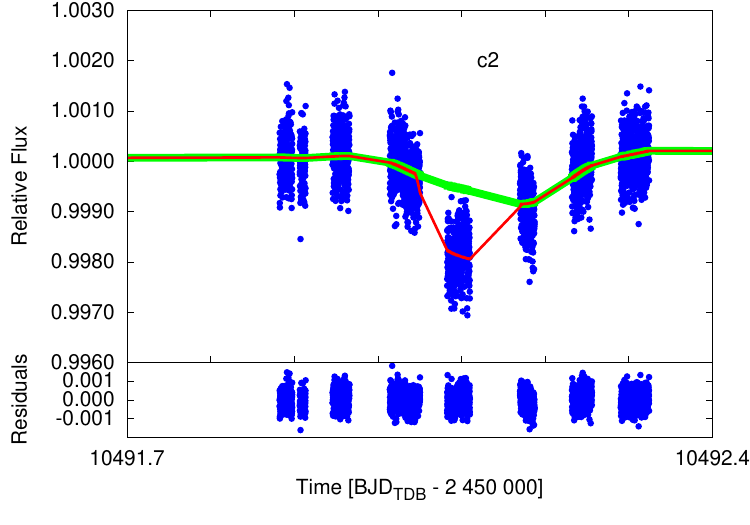}
\includegraphics[width=\textwidth/3]{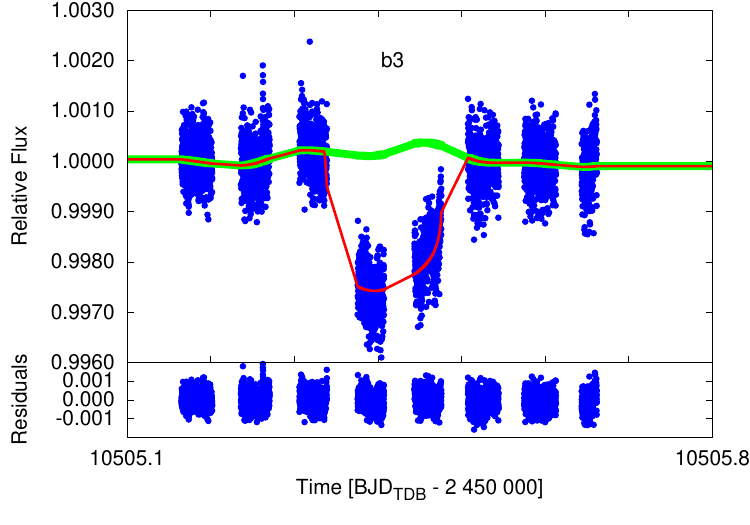}
\includegraphics[width=\textwidth/3]{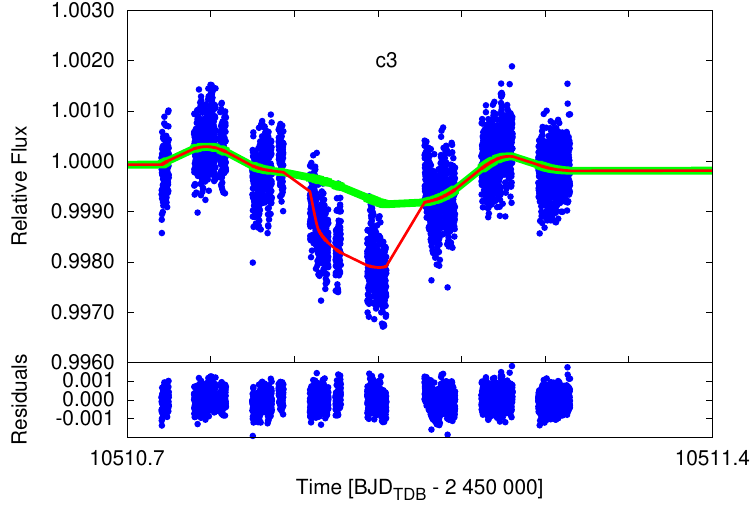}}
\centerline{
\includegraphics[width=\textwidth/3]{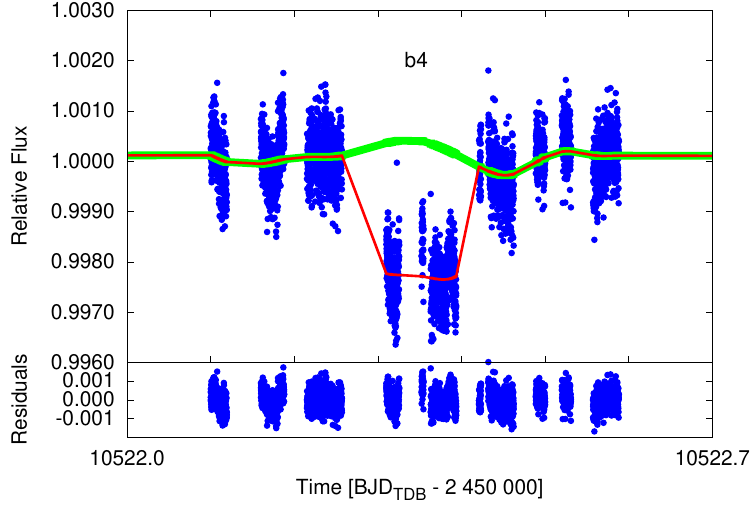}
\includegraphics[width=\textwidth/3]{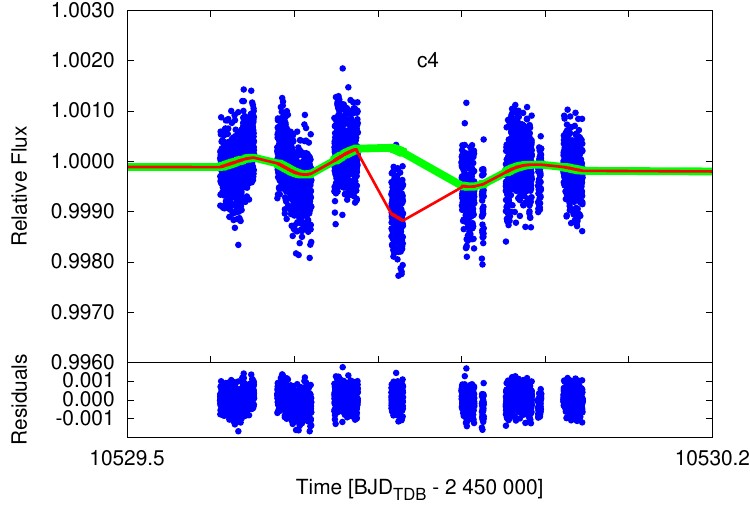}
\includegraphics[width=\textwidth/3]{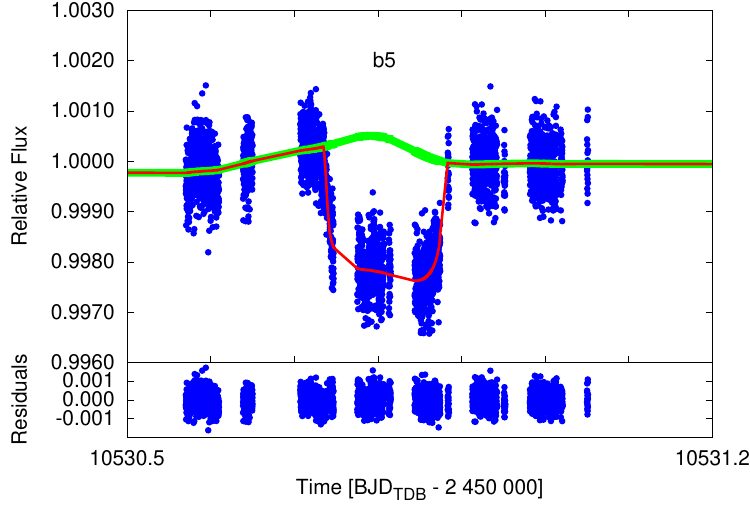}}
\centerline{
\includegraphics[width=\textwidth/3]{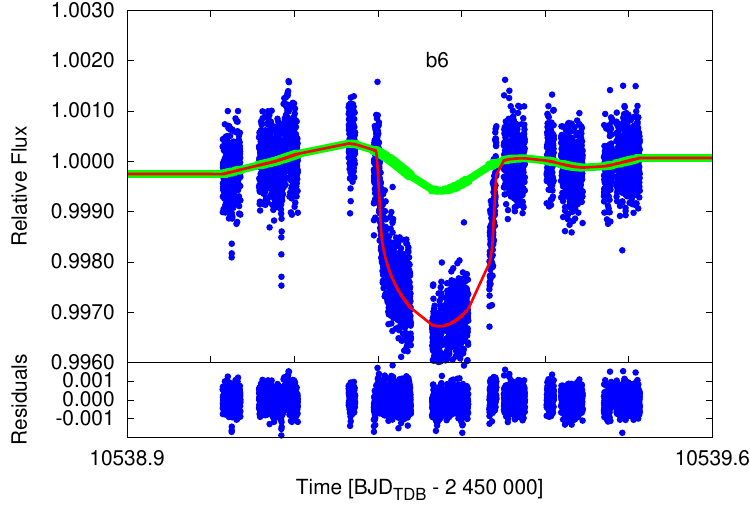}
\includegraphics[width=\textwidth/3]{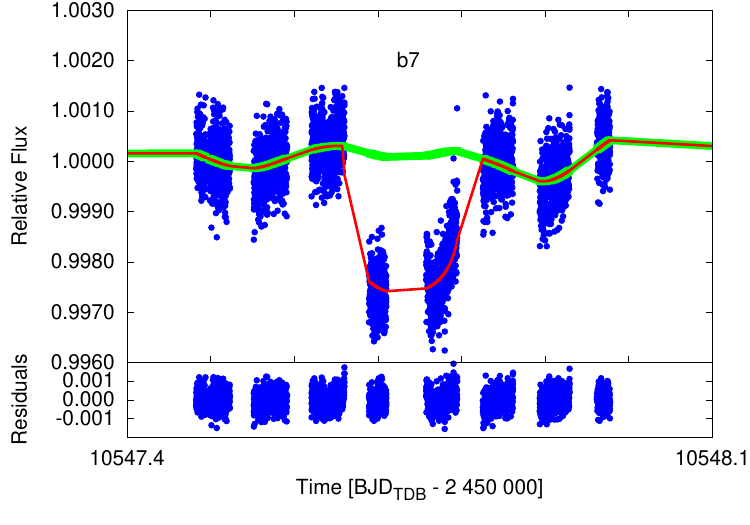}
\includegraphics[width=\textwidth/3]{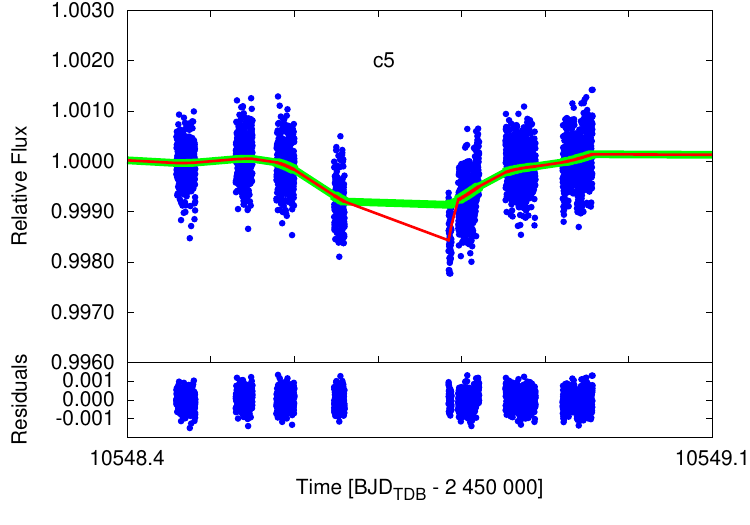}}
\centerline{
\includegraphics[width=\textwidth/3]{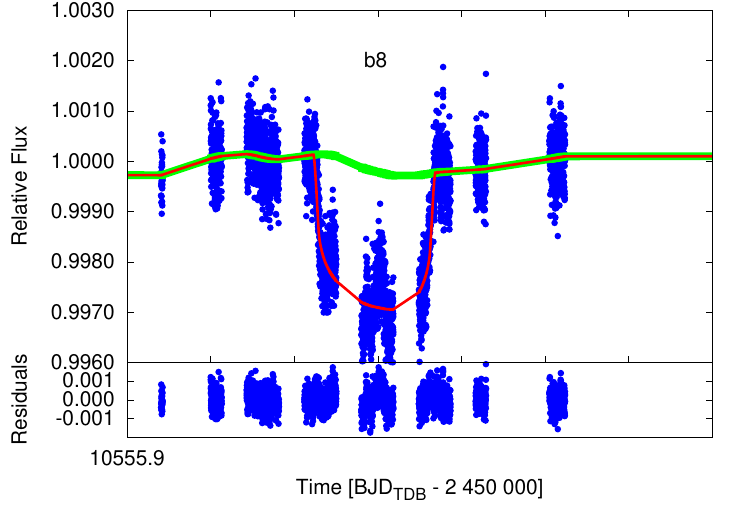}
\includegraphics[width=\textwidth/3]{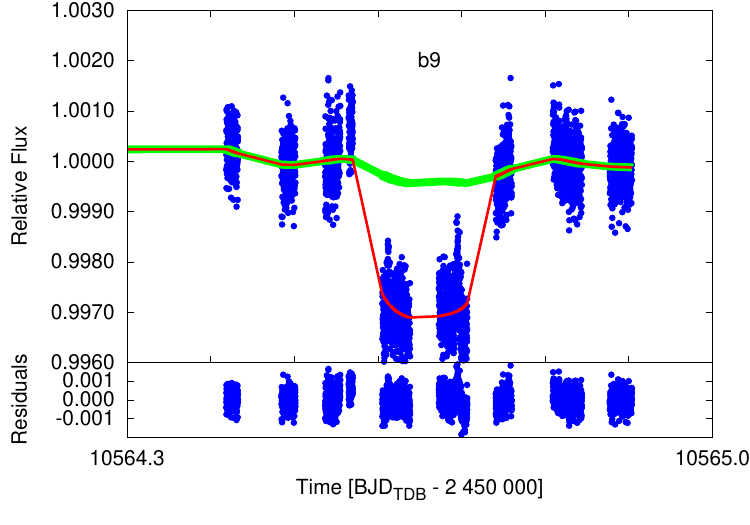}}
\caption{CHEOPS transit light curves of AU\,Mic\,b and AU\,Mic\,c from 2024, with the best-fitting \texttt{Allesfitter} full model. The blue data points are individual CHEOPS observations, the thick green line depicts the baseline model, and the red line represents the full model. Residuals are also shown. Visit numbers are indicated for easier identification of the observation.}
\label{fig:AU_Mic_planetary_transits_joint1}}
\end{figure}
\end{@twocolumnfalse}
\clearpage

\begin{@twocolumnfalse}

\begin{figure}[h]
\parbox{\textwidth}{
\centering
\centerline{
\includegraphics[width=\textwidth/3]{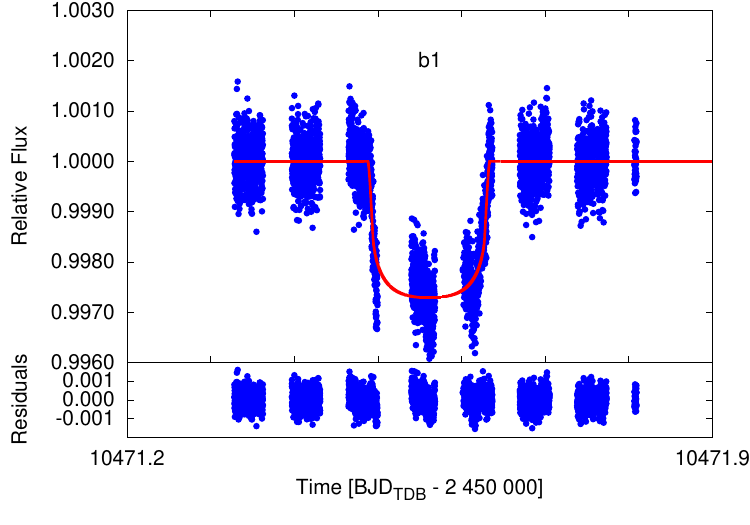}
\includegraphics[width=\textwidth/3]{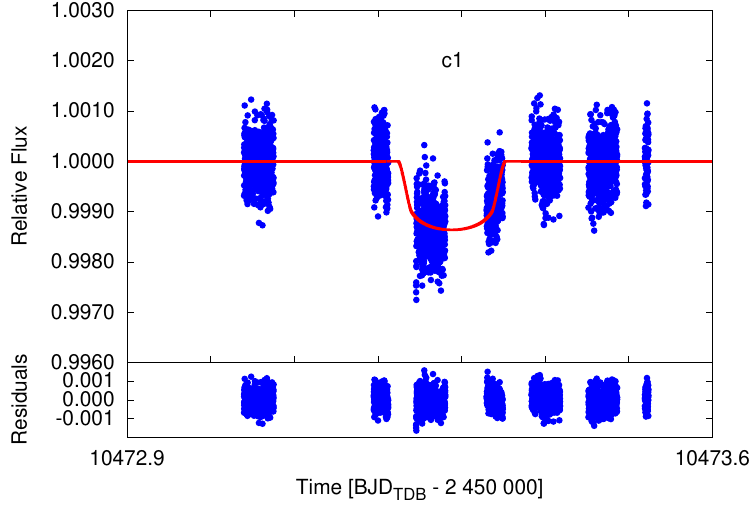}
\includegraphics[width=\textwidth/3]{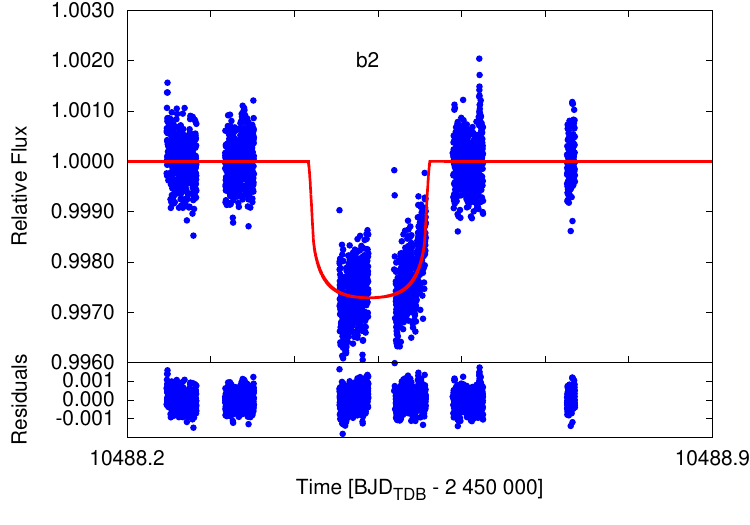}}
\centerline{
\includegraphics[width=\textwidth/3]{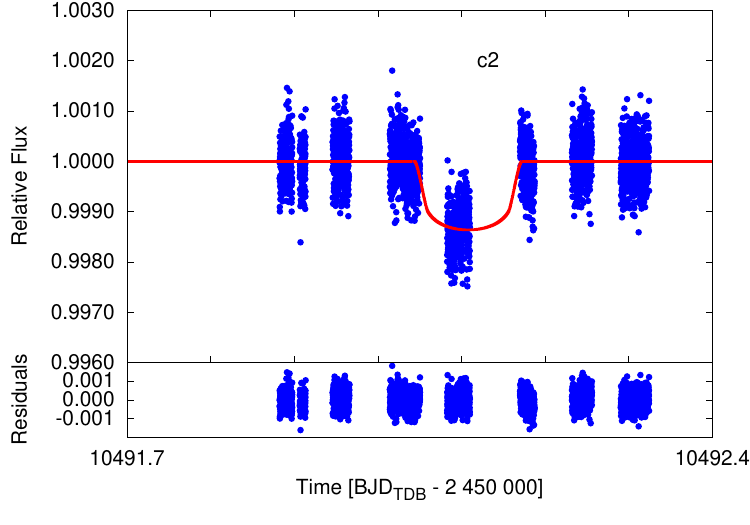}
\includegraphics[width=\textwidth/3]{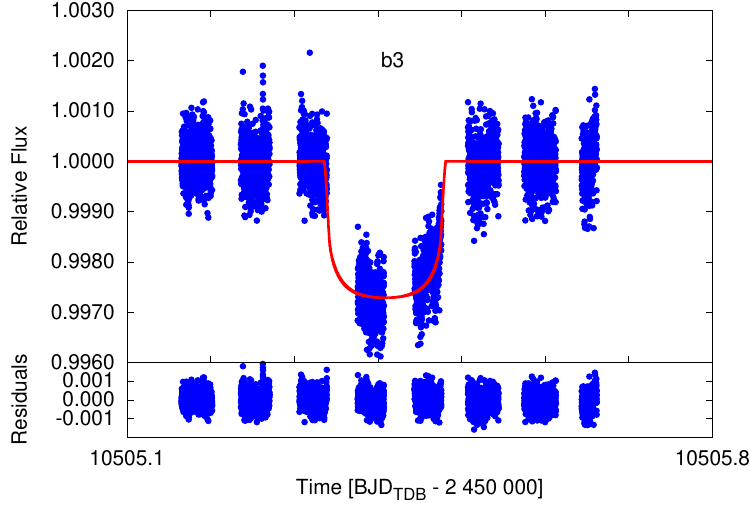}
\includegraphics[width=\textwidth/3]{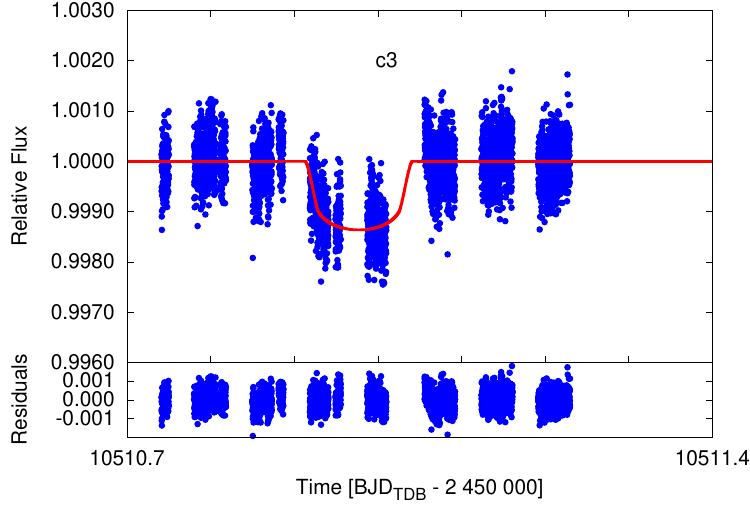}}
\centerline{
\includegraphics[width=\textwidth/3]{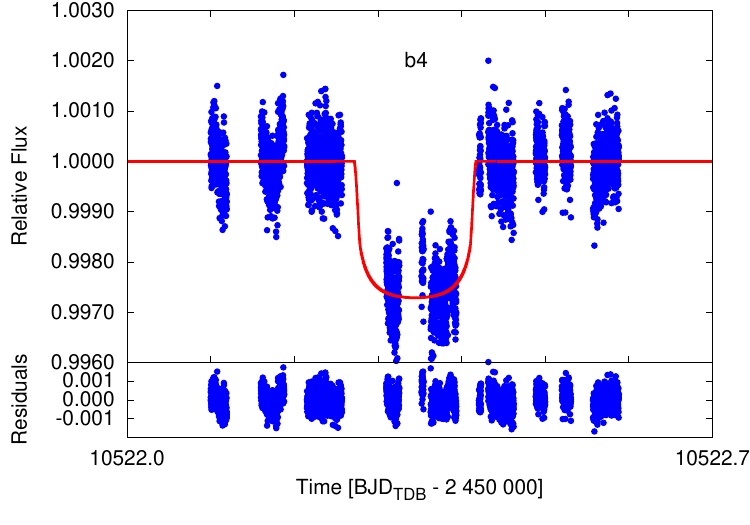}
\includegraphics[width=\textwidth/3]{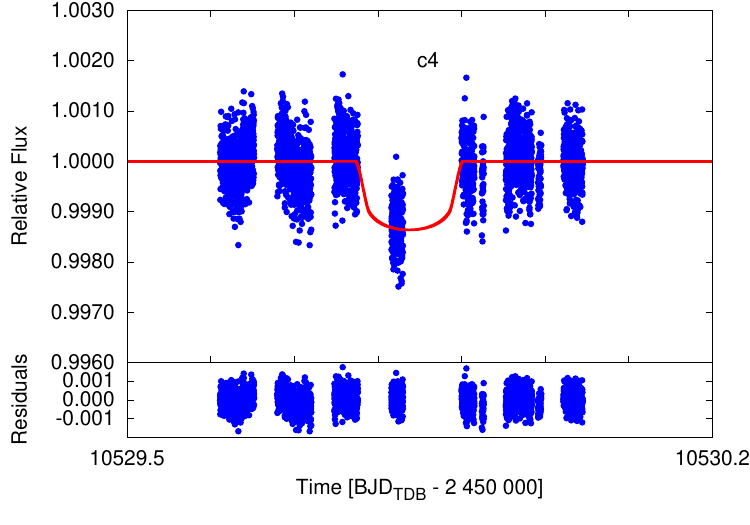}
\includegraphics[width=\textwidth/3]{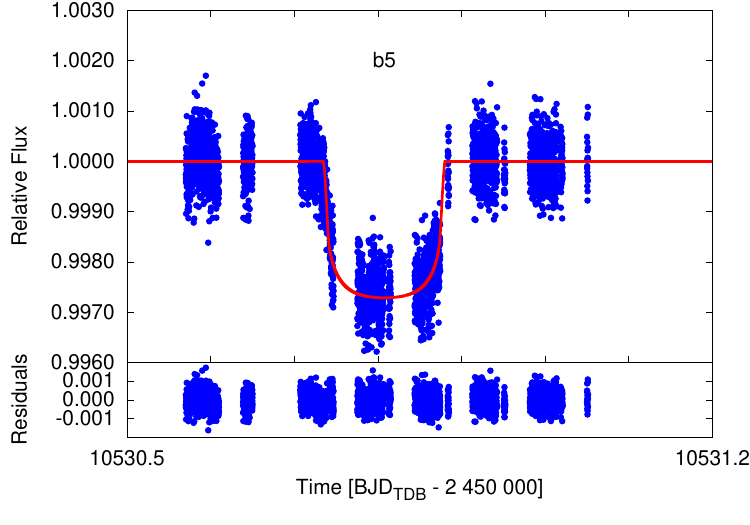}}
\centerline{
\includegraphics[width=\textwidth/3]{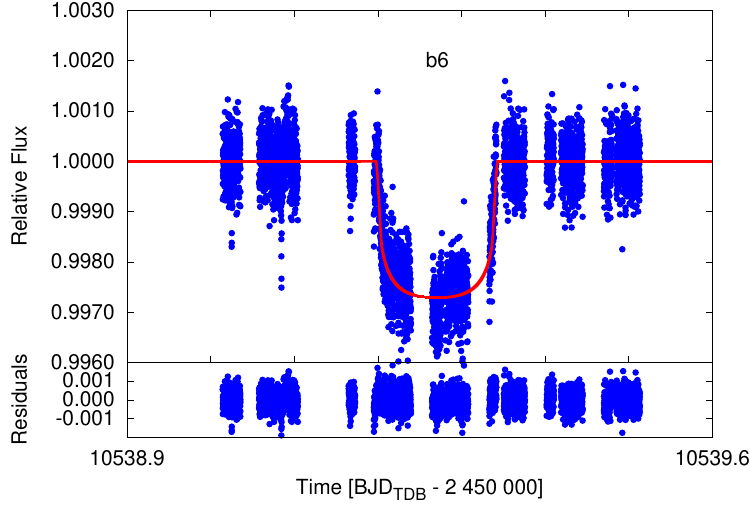}
\includegraphics[width=\textwidth/3]{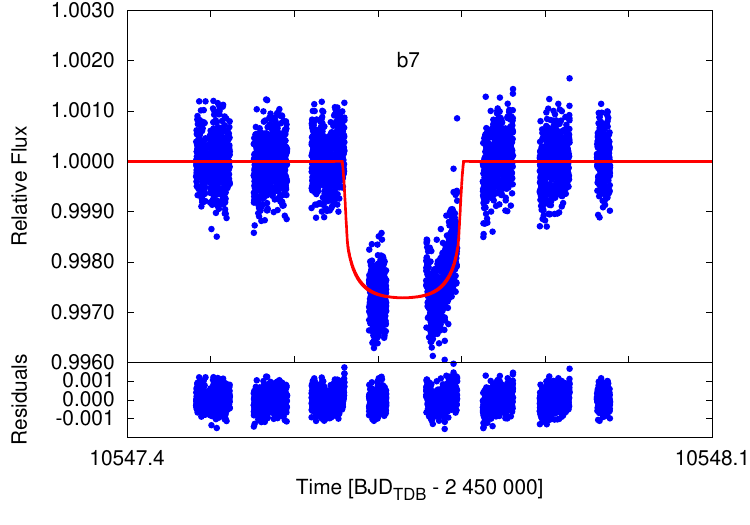}
\includegraphics[width=\textwidth/3]{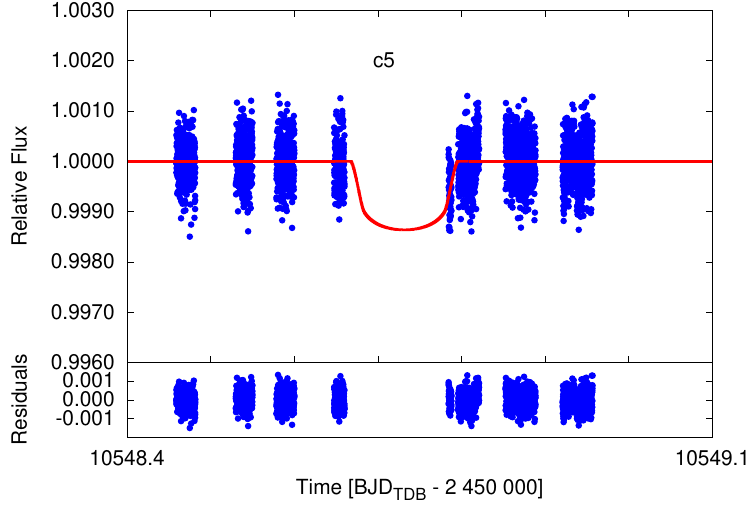}}
\centerline{
\includegraphics[width=\textwidth/3]{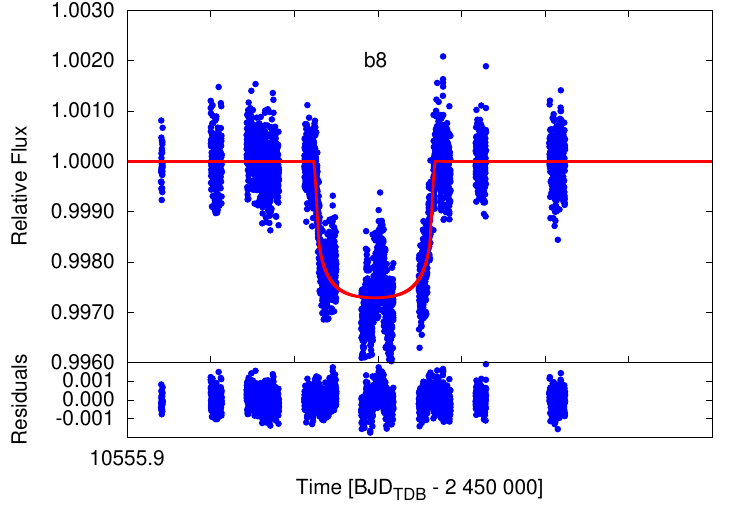}
\includegraphics[width=\textwidth/3]{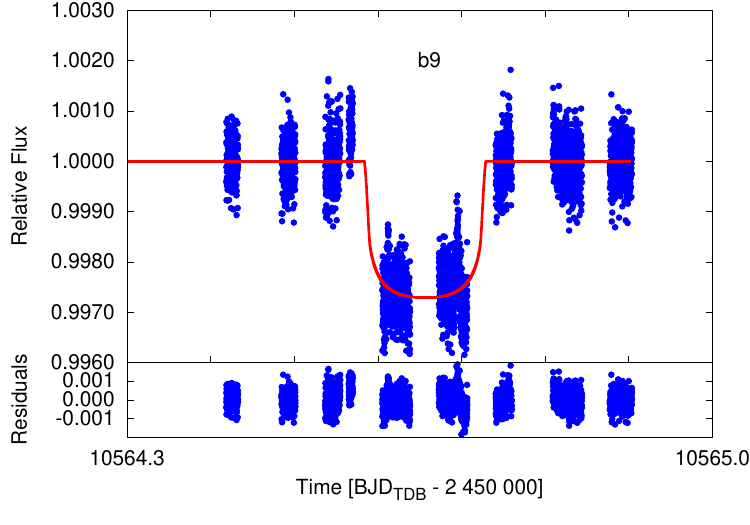}}
\caption{As in Fig. \ref{fig:AU_Mic_planetary_transits_joint1}, but only for the best-fitting \texttt{Allesfitter} transit model, which was over-sampled by a factor of 5 to better visualize the transit shape.}
\label{fig:AU_Mic_planetary_transits_joint2}}
\end{figure}

\clearpage

\begin{figure}[h]
\parbox{\textwidth}{
\centering
\centerline{
\includegraphics[width=\columnwidth]{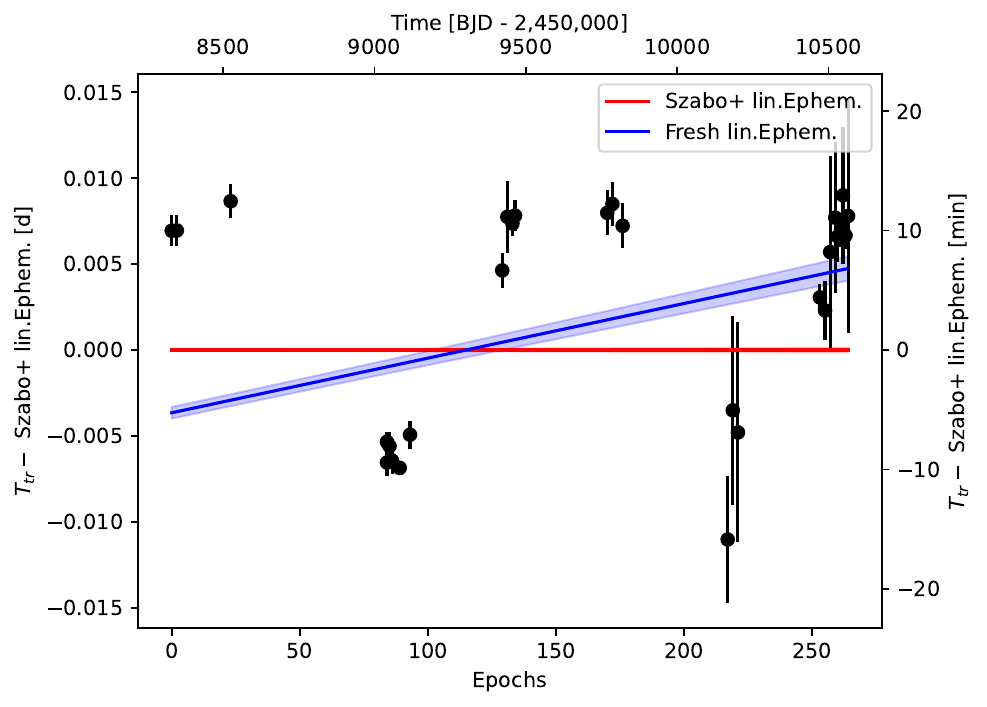}
\includegraphics[width=\columnwidth]{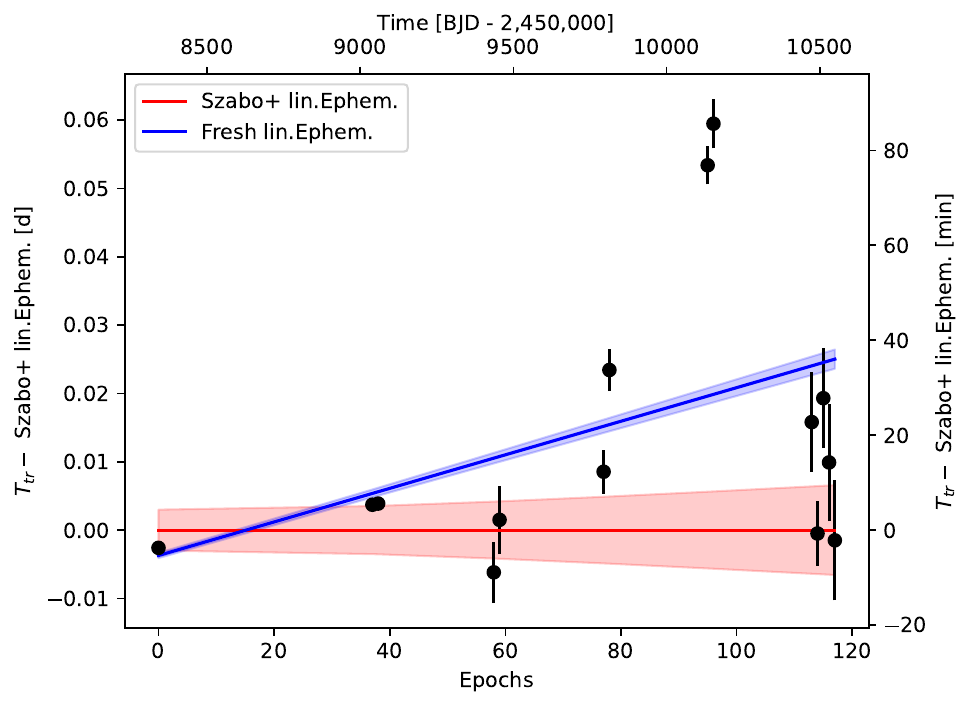}}
\centerline{
\includegraphics[width=\columnwidth]{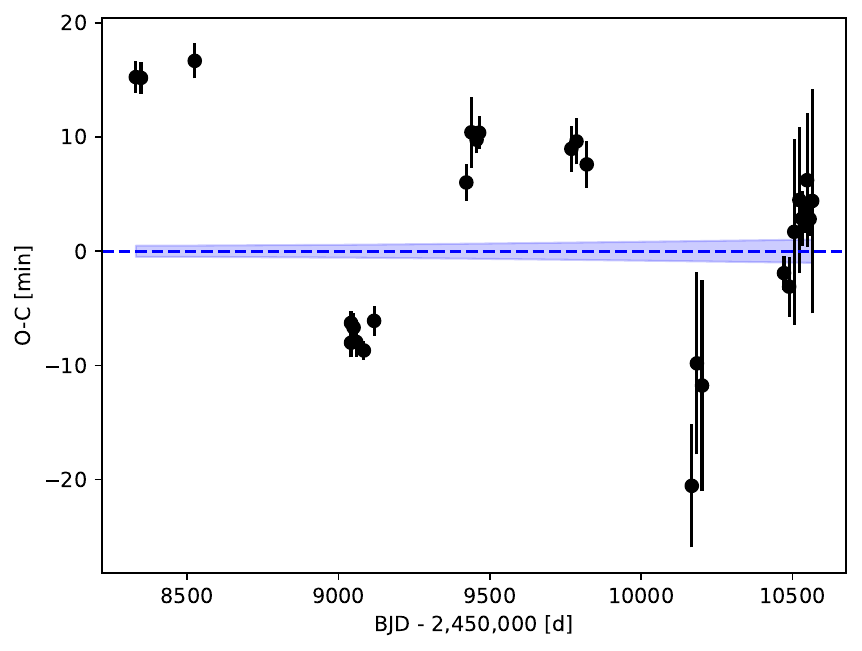}
\includegraphics[width=\columnwidth]{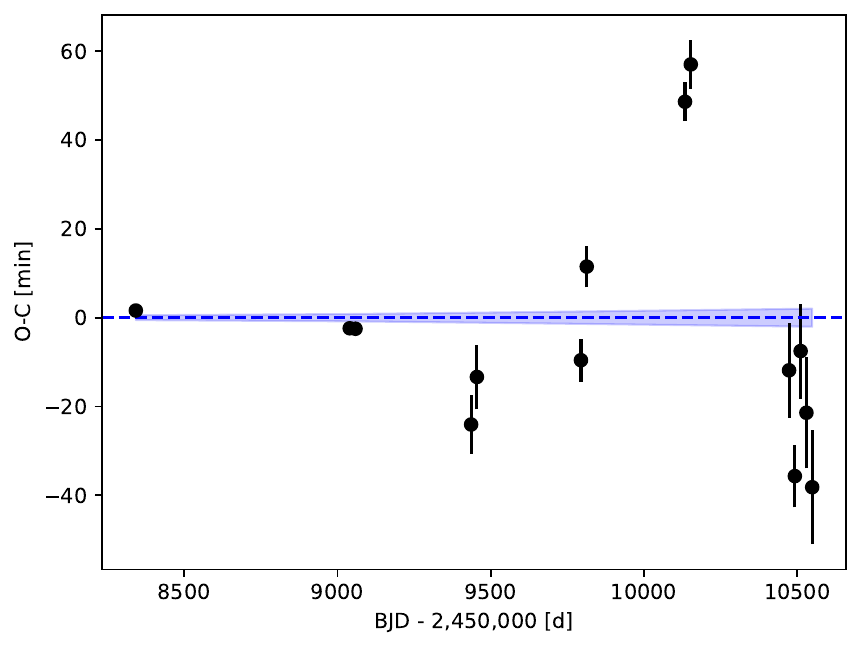}}
\caption{Comparison of the linear ephemerides' model from \citet{Szabo2} and from this work for AU\,Mic\,b (top left panel) and AU\,Mic\,c (top right panel). Observed-minus-calculated ($O-C$) diagrams of mid-transit times for AU\,Mic\,b (bottom left panel) and AU\,Mic\,c (bottom right panel), computed using the updated linear ephemerides from this work $T_\mathrm{c} = 2458330.38050 \pm 0.00034\,\mathrm{BJD_{TDB}}$, $P_\mathrm{orb} = 8.4631745 \pm 0.0000023\,\mathrm{d}$ (planet\,b), and $T_\mathrm{c} = 2458342.22320 \pm 0.00037\,\mathrm{BJD_{TDB}}$, $P_\mathrm{orb} = 18.859065 \pm 0.000011\,\mathrm{d}$ (planet\,c). The data points with uncertainties are from Tables\,\ref{table:ttvsb} and \ref{table:ttvsc}. The shaded areas mark the $1\sigma$ uncertainty region of the linear ephemerides’ model.}
\label{fig:AU_Mic_ttv}}
\end{figure}
\end{@twocolumnfalse}

\clearpage

\FloatBarrier

\section{Additional tables}
\label{app:tables}

This appendix contains additional tables supporting the analyses presented in the main text.  

\begin{@twocolumnfalse}

\begin{table}[h!]
\parbox{\textwidth}{
\centering
\caption{Priors and best-fitting parameters of AU\,Mic\,b and AU\,Mic\,c from \texttt{OCFit}.}
\label{ocfit-parameters-tab}
\begin{tabular}{lll}
\noalign{\smallskip}
\hline
\hline
\noalign{\smallskip}
Parameter & Prior & Value\\
\noalign{\smallskip}
\hline
\noalign{\smallskip}
\multicolumn{3}{c}{AU\,Mic\,b}\\
$a \sin i_3$ [au]                  					 & $\mathcal{U}$(0.0, 15.0) 		 		        & $1.22 \pm 0.04$\\
$e_3$ 		    								     & $\mathcal{U}$(0.0, 1.0)		 		            & $0.018 \pm 0.017$\\
$\omega_3$ [deg]                                     & $\mathcal{U}$(0.0, 360.0)                        & $54.5 \pm 2.6$\\
$T_\mathrm{c,3}$ [$\mathrm{BJD}_\mathrm{TDB}$]       & $\mathcal{U}$(2459470.0, 2459480.0)              & $2459475.0 \pm 2.8$\\
$P_\mathrm{mod,b}$ [d]                               & $\mathcal{U}$(1130.0, 1170.0)                    & $1167.6 \pm 20.3$\\
\noalign{\smallskip}
\hline
\noalign{\smallskip}
\multicolumn{3}{c}{AU\,Mic\,c}\\
$a \sin i_3$ [au]                  					 & $\mathcal{U}$(0.0, 15.0) 		 		        & $8.3 \pm 4.2$\\
$e_3$ 		    								     & $\mathcal{U}$(0.0, 1.0)		 		            & $0.934 \pm 0.071$\\
$\omega_3$ [deg]                                     & $\mathcal{U}$(0.0, 360.0)                        & $143 \pm 78$\\
$T_\mathrm{c,3}$ [$\mathrm{BJD}_\mathrm{TDB}$]       & $\mathcal{U}$(2458000.0, 2461000.0)              & $2460310 \pm 260$\\
$P_\mathrm{mod,c}$ [d]                               & $\mathcal{U}$(2000.0, 20000.0)                   & $2150 \pm 110$\\
\noalign{\smallskip}
\hline
\noalign{\smallskip}
\end{tabular}}
\end{table}

\begin{table}[h!]
\parbox{\textwidth}{
\centering
\caption{Selected parameters of AU\,Mic\,b compared to the results from previous works.}
\label{tab:solutionsb}
\begin{tabular}{ccccccc}
\noalign{\smallskip}
\hline
\hline
\noalign{\smallskip}
Dataset & $R_\mathrm{p}/R_\mathrm{s}$ & $a/R_\mathrm{s}$ & $R_\mathrm{p}$ [$R_{\oplus}$] & $a$ [au] & $b$ & Reference\\
\noalign{\smallskip}
\hline
\noalign{\smallskip}
2024 & $0.04811_{-0.00022}^{+0.00028}$ & $18.46_{-0.39}^{+0.47}$ & $4.31 \pm 0.11$ & $0.0705_{-0.0023}^{+0.0024}$ & $0.153_{-0.069}^{+0.100}$ & This work\\
2023 & $0.0517 \pm 0.0011$ & $18.18_{-0.42}^{+0.51}$ & $4.62 \pm 0.15$ & $0.0694_{-0.0023}^{+0.0025}$ & $0.386_{-0.074}^{+0.057}$ & \citet{Boldog1}\\
2022 & $0.04700_{-0.00073}^{+0.00077}$ & $18.83_{-0.54}^{+0.47}$ & $4.20 \pm 0.12$ & $0.0717 \pm 0.0026$ & $0.290_{-0.098}^{+0.077}$ & \citet{Boldog1}\\
2018--2021 & $0.0499 \pm 0.0004$ & $17.51_{-1.24}^{+1.12}$ & $4.79 \pm 0.29$ & $0.070_{-0.007}^{+0.006}$ & $0.502_{-0.048}^{+0.044}$ & \citet{Mallorquin1}\\
2018--2021 & $0.0488 \pm 0.0010$ & $18.79_{-0.59}^{+0.50}$ & $3.96 \pm 0.15$ & $0.0649 \pm 0.0012$ & $0.134_{-0.089}^{+0.096}$ & \citet{Wittrock2}\\
2021 & $0.0433 \pm 0.0017$ & $18.95 \pm 0.35$ & $3.55 \pm 0.13$ & $0.0654 \pm 0.0012$ & $0.17 \pm 0.11$ & \citet{Szabo2}\\
2020 & $0.0531 \pm 0.0023$ & $19.24 \pm 0.37$ & $4.36 \pm 0.18$ & $0.0678 \pm 0.0013$ & $0.09 \pm 0.05$ & \citet{Szabo1}\\
2018--2020 & $0.0512 \pm 0.0020$ & $18.5_{-1.4}^{+1.3}$ & $4.19_{-0.22}^{+0.27}$ & $0.0644_{-0.0054}^{+0.0056}$ & $0.26_{-0.17}^{+0.13}$ & \citet{Gilbert1}\\
2018--2020 & $0.0526_{-0.0002}^{+0.0003}$ & $19.1_{-0.4}^{+0.2}$ & $4.07 \pm 0.17$ & $0.0645 \pm 0.0013$ & $0.18 \pm 0.11$ & \citet{Martioli1}\\
2018 & $0.0514 \pm 0.0013$ & $19.1_{-1.6}^{+1.8}$ & $4.29 \pm 0.20$ & $0.066_{-0.006}^{+0.007}$ & $0.16_{-0.11}^{+0.14}$ & \citet{Plavchan1}\\
\noalign{\smallskip}
\hline
\noalign{\smallskip}
\end{tabular}}
\end{table}

\begin{table}[h!]
\parbox{\textwidth}{
\centering
\caption{Selected parameters of AU\,Mic\,c compared to the results from previous works.}
\label{tab:solutionsc}
\begin{tabular}{ccccccc}
\noalign{\smallskip}
\hline
\hline
\noalign{\smallskip}
Dataset & $R_\mathrm{p}/R_\mathrm{s}$ & $a/R_\mathrm{s}$ & $R_\mathrm{p}$ [$R_{\oplus}$] & $a$ [au] & $b$ & Reference\\
\noalign{\smallskip}
\hline
\noalign{\smallskip}
2024 & $0.03828_{-0.00050}^{+0.00045}$ & $29.60_{-0.49}^{+0.63}$ & $3.421 \pm 0.095$ & $0.1130_{-0.0034}^{+0.0035}$ & $0.8526 \pm 0.0097$ & This work\\
2023 & $0.0309_{-0.0033}^{+0.0028}$ & $30.68 \pm 0.96$ & $2.76_{-0.29}^{+0.26}$ & $0.1169 \pm 0.0047$ & $0.658_{-0.23}^{+0.093}$ & \citet{Boldog1}\\
2022 & $0.0354 \pm 0.0016$ & $30.7 \pm 1.0$ & $3.17 \pm 0.16$ & $0.1169 \pm 0.0047$ & $0.615_{-0.11}^{+0.074}$ & \citet{Boldog1}\\
2018--2021 & $0.0291_{-0.0003}^{+0.0005}$ & $29.87_{-2.11}^{+1.91}$ & $2.79_{-0.17}^{+0.18}$ & $0.119 \pm 0.011$ & $0.259_{-0.178}^{+0.203}$ & \citet{Mallorquin1}\\
2018--2021 & $0.0311 \pm 0.0028$ & $32.05_{-1.00}^{+0.86}$ & $2.52 \pm 0.25$ & $0.1108 \pm 0.0020$ & $0.30_{-0.19}^{+0.13}$ & \citet{Wittrock2}\\
2021 & $0.0313 \pm 0.0016$ & $28.8 \pm 2.4$ & $2.56 \pm 0.12$ & $0.0993 \pm 0.0085$ & $0.58 \pm 0.13$ & \citet{Szabo2}\\
2018--2020 & $0.0340_{-0.0033}^{+0.0034}$ & $31.7_{-2.7}^{+2.6}$ & $2.79_{-0.30}^{+0.31}$ & $0.110 \pm 0.010$ & $0.30_{-0.20}^{+0.21}$ & \citet{Gilbert1}\\
2018--2020 & $0.0418_{-0.0012}^{+0.0010}$ & $29 \pm 3.0$ & $3.24 \pm 0.16$ & $0.1101 \pm 0.0022$ & $0.51 \pm 0.21$ & \citet{Martioli1}\\
\noalign{\smallskip}
\hline
\noalign{\smallskip}
\end{tabular}}
\end{table}
\end{@twocolumnfalse}

\FloatBarrier

\end{appendix}

\end{document}